\documentclass[12pt]{article}

\newsavebox{\ns}
\newsavebox{\dbrane}
\newsavebox{\dbshort}

\usepackage{epsfig}
\usepackage{amssymb}

\def\be{\begin{eqnarray}}
\def\ee{\end{eqnarray}}

\newcommand{\nn}{\nonumber}
\newcommand\para{\paragraph{}}
\newcommand{\ft}[2]{{\textstyle\frac{#1}{#2}}}
\newcommand{\eqn}[1]{(\ref{#1})}

\def\Dslash{\,\,{\raise.15ex\hbox{/}\mkern-12mu D}}
\def\Dbarslash{\,\,{\raise.15ex\hbox{/}\mkern-12mu {\bar D}}}
\def\delslash{\,\,{\raise.15ex\hbox{/}\mkern-9mu \partial}}
\def\delbarslash{\,\,{\raise.15ex\hbox{/}\mkern-9mu {\bar\partial}}}
\def\pslash{\,\,{\raise.15ex\hbox{/}\mkern-9mu p}}
\def\calDslash{\,\,{\raise.15ex\hbox{/}\mkern-12mu {\cal D}}}

\newcommand{\sign}{{\rm sign}}
\newcommand{\tr}{{\rm tr}}
\newcommand{\Tr}{{\rm Tr}}
\newcommand{\valpha}{{\vec{\alpha}}}
\newcommand{\vbeta}{{\vec{\beta}}}

\begin{document}
\pagestyle{plain}
\setcounter{page}{1}
\newcounter{bean}
\baselineskip16pt

\begin{titlepage}

\begin{flushright}
TIT/HEP--534 \\
{\tt hep-th/0501207} \\
January, 2005 \\
\end{flushright}
\vspace{3mm}

\begin{center}
{\ \ }

\vskip 1.0 cm
{\large \bf Monopoles, Vortices, Domain Walls and D-Branes:}
%: The Rules of Interaction}
\vskip 0.15cm
{\large\bf The Rules of Interaction}
\vskip 1 cm
{Norisuke Sakai${}^1$ and David Tong${}^2$}\\
\vskip 1cm
${}^1${\sl Department of Physics, Tokyo Institute of Technology, \\
Tokyo  152-8551, Japan \\}
{\tt nsakai@th.phys.titech.ac.jp}
\vskip 0.4 cm
${}^2${\sl Department of Applied Mathematics and Theoretical Physics, \\
University of Cambridge, CB3 OWA, UK \\}
{\tt d.tong@damtp.cam.ac.uk}

\end{center}

\vskip 0.5 cm
\begin{abstract}
Non-abelian gauge theories in the Higgs phase admit a 
startling variety of BPS solitons. These include domain 
walls, vortex strings, confined monopoles threaded on vortex 
strings, vortex strings ending on domain walls, 
monopoles threaded on strings ending on domain walls, and more. 
After presenting a self-contained review of these objects, 
including several new results on the dynamics of domain walls, 
we go on to examine the possible interactions of solitons of 
various types. We point out the existence of a classical 
binding energy when the string ends on the domain wall which 
can be thought of as a BPS boojum with negative mass. 
We present an index theorem for domain walls in non-abelian 
gauge theories. 
We also answer questions such as: 
Which strings can end on which walls? What happens when 
monopoles pass through domain walls? What happens when domain 
walls pass through each other? 
\end{abstract}

\end{titlepage}

\section{Introduction}
\label{sc:introduction}

Non-abelian gauge theories in the phase where the gauge group 
is fully broken enjoy a wonderfully rich spectrum of solitons. 
In this paper we review some of the recent results on these 
objects and specify the allowed rules of interaction 
between the solitons of different types.

\para
In Section \ref{sc:theory}  
we describe the theory in question: it is a $d=3+1$ dimensional 
$U(N_c)$ gauge theory with $N_f$ Higgs fields in the fundamental 
representation and a single scalar field in the adjoint 
representation. These fields comprise the bosonic sector 
of ${\cal N}=2$ supersymmetric QCD and all the solitons 
described in this paper are BPS in nature, preserving some fraction 
of the supersymmetry. After examining 
the symmetry and vacuum structure of the theory, the remaining 
sections are devoted to studying each 
of the various solitons in turn. They are:

\para
{\bf Domain Walls:} As we shall explain in 
Section \ref{sc:theory}, the 
theory has a large number of isolated vacuum states. 
In fact, this number grows exponentially 
as the number of flavours is increased. 
This allows for a vast selection of co-dimension one domain 
wall solutions interpolating between different vacua and 
raises various questions: what is the correct way to classify 
the domain walls? Given two vacua, how many 
domain walls can interpolate between them? What properties do 
these domain walls have? In Section \ref{sc:domain-wall} 
we shall review what is known about these domain walls and present 
answers to all these questions. An appendix contains an index theorem 
in which we determine the number of parameters of a given domain wall 
system.

\para
{\bf Vortices:} The theory also admits co-dimension two solitonic objects.
These are vortex strings. In fact, in any
given vacuum there are $N_c$ different types of vortex strings corresponding to
the magnetic field sitting in a different diagonal component of
the adjoint-valued field strength. In 
Section \ref{sc:vortices} we review recent
work on the interactions of these different vortex strings.

\para
{\bf Confined Monopoles:} Famously, non-abelian gauge theories which
are broken to $U(1)$ factors admit 't Hooft-Polyakov monopoles.
Recently it was realised that stable monopoles continue to exist in
the theory even when the gauge group is broken completely. Moreover,
these monopoles
remain BPS. Rather than spreading out radially,
the magnetic flux leaves the monopole in two collimated 
tubes which are identified with the vortex strings. 
From the perspective of the vortex, the monopole looks 
like a kink solution in the string. We 
review this object in Section \ref{sc:confined-monopoles}.

\para
{\bf D-Branes:} The domain walls of our theory have an interesting
property: a $U(1)$ gauge field lives on their $d=2+1$ dimensional
worldvolume. Moreover, the vortex string can terminate
on the domain wall where its end is electrically charged under
the $U(1)$ gauge field. In other words, the theory contains a
semi-classical D-brane configuration. We describe this soliton
in Section \ref{sc:d-branes} %6 
and show that there exists a property of this system
missed in previous studies: a finite, classical binding energy between
the end of the string and the domain wall. This binding energy
can be thought of as a BPS, negative mass, magnetic monopole - or boojum - 
embedded within the domain wall.

\para
{\bf Interactions:}
Given these basic objects, it is natural to study the dynamics of
the various components. For a single class of solitons -- such
as domain walls, or vortices -- this is a well developed
subject. But less attention has been paid to the dynamics of
composite
configurations that include multiple types of solitons. 
In Sections \ref{sc:d-branes} %6 
and 7, we describe the basic rules of classical 
interaction between solitons of different types. 
We answer questions such as: Which type of vortex can 
end on which type of domain wall? Can monopoles pass through domain walls?
How do monopoles interact with the string binding energy? What becomes of
strings and monopoles when domain walls pass through each other?

%\para
%We include two appendices. Appendix A describes the relationship 
%between these gauge theory solitons and solitons in massive 
%sigma-models. In Appendix B we present the index theorem for 
%domain walls in non-abelian gauge theories. 

\section{The Theory}
\label{sc:theory}

Throughout this paper we shall consider a class of $d=3+1$ 
dimensional\footnote{All solutions described here can be uplifted 
to $d=4+1$ dimensions where supersymmetry allows real masses for 
hypermultiplets.} non-abelian gauge theories with ${\cal N}=2$ supersymmetry. 
We take a $U(N_c)$ gauge group, coupled to $N_f$ 
hypermultiplets transforming in the fundamental representation. 
Although the solitons exist most naturally in this 
supersymmetric context, for the most part we need only 
discuss the bosonic fields and we focus on these, 
commenting on the fermionic zero modes only in passing.

\para
The $U(N_c)$ vector multiplet includes the gauge field 
$A_\mu$ and a complex adjoint scalar field $\phi$, together 
with fermions. 
The bosonic content of each hypermultiplet consists of two 
complex scalar fields $q_i$ and $\tilde{q}_i$, where 
$i=1,\ldots, N_f$ is the flavour index. 
Each field $q$ transforms in the fundamental ${\bf N_c}$ 
representation of the gauge group, while $\tilde{q}$ 
transforms in the anti-fundamental ${\bf{\bar N}_c}$ 
representation. 
Denoting spacetime indices by $M, N %\mu, \nu
=0,1,2,3$, we obtain 
the bosonic part of the Lagrangian\footnote{
Our convention for the spacetime metric is 
$\eta_{MN%\mu\nu
}={\rm diag}(-1,+1,+1,+1)$. 
We represent the fields in the vector multiplet in terms of 
matrices in the fundamental representation of $U(N_c)$ 
whose generators $T^a$ are normalized as 
Tr$(T^a T^b)=\delta^{ab}$. } 
\be
-{\cal L}
&=&\frac{1}{2e^2}{\Tr} \left(\ft12 F_{MN%\mu\nu
}F^{MN%\mu\nu
}
+|{\cal D}_M%\mu
\phi|^2
+|[\phi,\phi^\dagger]|^2 \right)
+\sum_{i=1}^{N_f}\left(|{\cal D}_M%\mu
 q_i|^2
+|{\cal D}_M%\mu
\tilde{q}_i|^2 \right)
\nn\\ && +
{\Tr}\left(\frac{e^2}{2}(\sum_{i=1}^{N_f}(q_iq^{\dagger}_i
-\tilde{q}^\dagger_i\tilde{q}_i)-v^2)^2
+e^2|\sum_{i=1}^{N_f}\tilde{q}_iq_i|^2\right)
\label{lag}\\ &&
+ \frac{1}{2}\left(\sum_{i=1}^{N_f}(q_i^\dagger\left\{\phi-m_i,
\phi^\dagger-m_i^\dagger\right\}
q_i + \tilde{q}_i\left\{\phi-m_i,\phi^\dagger-m_i^\dagger
\right\}\tilde{q}_i^\dagger\right)
\nn\ee
Here the adjoint covariant derivative is given by 
${\cal D}_M%\mu
\phi=\partial_M%\mu
\phi-i[A_M%\mu
,\phi]$ while 
the fundamental covariant derivative is 
${\cal D}_M%\mu 
q=\partial_M%\mu 
q-iA_M%\mu 
q$. 
The field strength is defined by 
$F_{MN%\mu\nu
}=i[{\cal D}_M%\mu
,{\cal D}_N%\nu
]$. 

\para
The theory contains three types of parameters: 
the gauge coupling $e^2$, the Higgs vacuum expectation value 
(vev) (or, in the language of supersymmetry, the 
Fayet-Iliopoulos (FI) parameter) $v$, and the complex masses 
$m_i$. It will turn out that the soliton structure is richest 
when we set the masses to be real and distinct\footnote{This 
is entirely analogous to the situation with 't Hooft-Polyakov 
monopoles where the moduli spaces have largest dimension when 
only a single real adjoint scalar field has a vev.}: 
$m_i=m_i^\dagger\neq m_j$ for $i\neq j$. In particular, this 
allows us to order the masses 
$m_i>m_{i+1}$.

\para
The supersymmetric vacua of the theory exist when all terms 
in the potential can be set to zero. 
The presence of the vev $v$ means that the $q$ fields must be 
non-vanishing in order that the second line of \eqn{lag} 
vanishes. In turn, the presence of the masses $m_i$ then 
requires that the $\phi$ field is non-zero so that the third 
line in \eqn{lag} vanishes. 
The choice of a vacuum is equivalent to picking a set 
$\Sigma$ of $N_c$ distinct elements out 
of a possible $N_f$,
\be
\Sigma = \left\{\sigma(a): \sigma(a)\neq \sigma(b)\ {\rm for}\ a\neq b\right\}
\label{set}\ee
where $a=1,\ldots,N_c$ runs over the colour index, while
$\sigma(a)\in \{1,\ldots,N_f\}$. Up to a Weyl transformation, the vacuum is given by,
\be
\phi={\rm diag}(m_{\sigma(1)},\ldots,m_{\sigma(N_c)})
\ \ \ \ \ \ ,\ \ \ \  \ \ \
q^a_{\ i}=v\,\delta^a_{\ i=\sigma(a)}
\label{vacuum}\ee
supplemented by $\tilde{q}=0$ \cite{nr}. For $N_f<N_c$ there are no 
supersymmetric vacua; for $N_f\geq N_c$, 
the number of vacua is 
$N_{\rm vac}=\left(N_f{\rm -choose-} N_c \right)=N_f!/N_c!(N_f-N_c)!$. 
Each of these vacua is isolated and gapped. 
Parametrically, there are $N_c^2$ non-BPS massive gauge 
bosons and quarks with masses 
$m^2_\gamma\sim e^2v^2+|m_{\sigma(a)}-m_{\sigma(b)}|^2$ and 
$N_c(N_f-N_c)$ BPS massive quark fields with masses 
$m_q\sim|m_{\sigma(a)}-m_i|$ (with $i\notin \Sigma$). 

\para
The $U(N_c)$ gauge symmetry and the $SU(N_f)$ flavour 
symmetry of the theory are broken both explicitly by the 
masses $m_i$ and spontaneously by the vev $v^2$. 
In each of the vacua, the pattern of symmetry breaking is\footnote{The notation  
$S[\otimes_i U(N_i)]$ means we project out the diagonal, central $U(1)$ 
from $\otimes_i U(N_i)$.}
\be
U(N_c)\times SU(N_f)\longrightarrow 
S[\,U(1)^{N_c}_{\rm diag}\times U(1)^{N_f-N_c}\,]
\label{breaking}\ee
Since the surviving unbroken group involves a simultaneous 
gauge and flavour rotation, the theory is said to lie in 
the colour-flavour-locked phase. 
However, note that in different vacua a different 
$U(1)^{N_c}\subset U(1)^{N_f-1}\subset SU(N_f)$ is 
locked with the gauge group, the choice determined by 
\eqn{set}. 
This will play an important role when we discuss 
vortices connecting to domain walls. 
%
%
%\para
%The $U(N_c)$ gauge symmetry and $SU(N_f)$ flavour symmetry of the theory
%are broken both explicitly by the masses $m_i$ and spontaneously by the
%vev $v^2$. The pattern of symmetry breaking at intermediate scales
%depends on the relative values of these parameters. For $|m_i-m_j| \gg ev$,
%the flavour group is explicitly broken by masses at a higher spontaneous
%symmetry breaking by $v^2$. Since the adjoint scalar picks up a vev related
%to the mass \eqn{vacuum}, the symmetry breaking is
%\be
%U(N_c)_G\times SU(N_f)_F\stackrel{m}{\longrightarrow} U(1)_G^{N_c}\times U(1)_F^{N_f-1}
%\stackrel{v}{\longrightarrow}S[U(1)^{N_c}_{\rm diag}\times U(1)^{N_f-N_c}_F]
%\label{break1}\ee
%However, if $ev\gg |m_i-m_j|$, then the spontaneous breaking due to the vacuum
%expectation value of $q$ occurs at a higher scale than the explicit breaking due
%to masses,
%\be
%U(N_c)_G\times SU(N_f)_F\stackrel{v}{\longrightarrow} S[U(N_c)_{\rm diag}
%\times U(N_f-N_c)]\stackrel{m}{\longrightarrow} S[U(1)^{N_c}_{\rm diag}
%\times U(1)^{N_f-N_c}]
%\label{break2}\ee
%
%
%
\para
For the solutions we consider, most of these fields will 
not play a role and we shall set them to zero at this stage. 
The reality of the masses allows us to restrict to a real 
adjoint scalar field: $\phi={\rm Re}(\phi)$. 
We shall also set $\tilde{q}_i=0$. One can show that 
neither ${\rm Im}(\phi)$ nor
$\tilde{q}_i$ have zero modes on any of our soliton 
solutions. 
We shall further restrict to time independent solutions 
with vanishing electric fields $F_{0\alpha}=0$ although 
there do exist dyonic versions of many of the solitons 
discussed below in which this constraint is relaxed. 
Truncating to the surviving fields, the bosonic\footnote{
When including 
fermionic fields, it is crucial to keep both the 
superpartners of $q$ and $\tilde{q}$ to avoid anomalies. 
Moreover, both have zero modes on the soliton solutions 
described below.
} Hamiltonian  with which we shall work is, 
\be
{\cal H}= \frac{1}{2e^2}\Tr\left[B_\alpha^2+
({\cal D}_\alpha\phi)^2\right] 
+ \sum_{i=1}^{N_f}|{\cal D}_\alpha q_i|^2 
+ \sum_{i=1}^{N_f}q_i^\dagger(\phi-m_i)^2q_i
+\frac{e^2}{2}
\Tr%\left
[(\sum_{i=1}^{N_f}q_iq^\dagger_i-v^2)^2%\right
]
\label{ham}
\ee
Here $B_\alpha=\ft12\epsilon_{\alpha\beta\gamma}F_{\beta\gamma}$ 
is the non-abelian magnetic field, with 
$\alpha=1,2,3$ the spatial index. In the following sections 
we will embark on a tour of 
the wide variety of solutions on offer in this simple Hamiltonian.

\subsubsection*{\it Massive Gauged Linear Sigma-Models}

It is well known that the long-wavelength limit of certain supersymmetric
gauge theories is described by a sigma-model on the Higgs branch of
the theory which, in our case, is the cotangent bundle of the Grassmannian
$T^\star\,G(N_c,N_f)$.  This construction, often referred 
to as a gauged linear sigma 
model, has proven very useful in analysing non-linear sigma models, most
notably in the context of two-dimensional gauge theories \cite{phases}. 
In the present
case the masses for the scalar fields $q$ and $\tilde{q}$ lift the Higgs
branch. They can be thought of as inducing a potential on the sigma
model target space of a specific form \cite{agf}. The long-wavelength 
limit of our theory is therefore given by a massive sigma model. 
Many (although not all) of the soliton solutions we will 
describe below have analogs in these massive 
sigma models which correctly capture the coarser features of the gauge theory
configurations. Moreover,
the solitons are usually simpler in the sigma-model context (for example,
exact solutions can often be found in this limit) and, in  most cases,
were discovered there first. Since this connection between gauge theories
and sigma models is important for our solitons, in Appendix A we review how
the sigma model emerges from the classical $e^2\rightarrow\infty$ limit of
the gauge theory.

\section{Domain Walls}
\label{sc:domain-wall}

The existence of multiple, gapped, isolated vacua is sufficient to guarantee
the existence of co-dimension one domain walls. While domain walls
in supersymmetric Wess-Zumino models have been studied in detail,
the domain walls in the theory \eqn{lag} have received the attention they
deserve only recently. As we shall see, they have rather special
properties. They were first studied by Abraham and Townsend in \cite{at}.
Subsequently, many people have examined their various properties
in the sigma model limit \cite{intersect,multi,arai,losev}, in abelian
\cite{kinky,tong,keith,shif,junction,isozumi} and non-abelian \cite{shif2,allquarter,sakai}
gauge theories, and in quantum
theories \cite{nick,mirror}. In this section we review the domain wall
equations and present a new, simple method of classifying the domain walls
in terms of the root lattice of the flavour group. As we shall see, this
classification encodes much of the interesting information about the dynamics
of domain walls.

\para
Let us consider domain walls transverse to the $x^3$ direction. We
pick a vacuum $\Sigma$ at $x^3= +\infty$ as determined by a set
\eqn{set} and a distinct vacuum $\tilde{\Sigma}$ at $x^3=-\infty$. 
The first order domain wall equations
were first derived in \cite{kinky} and can be determined 
by the usual Bogomoln'yi ``completing-the-square'' trick. 
If we restrict to configurations with 
$\partial_1=\partial_2=A_1=A_2=0$, then the Hamiltonian 
\eqn{ham} can then be written as,
\be
\int dx^3\,{\cal H}&=&\int dx^3\ \frac{1}{2e^2}{\Tr}\,
[({\cal D}_3\phi \mp e^2 (\sum_i q_iq^\dagger_i-v^2))^2]
\pm {\Tr}\, %\left
[ ({\cal D}_3\phi)
(\sum_i q_iq_i^\dagger -v^2) %\right
] 
\nn\\&&
\ \ \ \ \ \ 
+ \sum_i \left( |{\cal D}_3q_i\mp (\phi-m_i)q_i|^2 
\pm q_i^\dagger (\phi-m_i)
{\cal D}_3q_i \pm {\cal D}_3 q_i^\dagger (\phi-m_i)q_i \right)
\nn\\
&\geq & \mp \left[\sum_i q_i^\dagger m_iq_i\right]_{-\infty}^{+\infty}
\pm\left[{\Tr}\ \phi(\sum_iq_iq_i^\dagger-v^2)\right]_{-\infty}^{+\infty}
\nn\\
&=&  \mp v^2\,\Tr[\phi]^{+\infty}_{-\infty}=
\mp v^2\sum_{a=1}^{N_c}(m_{\sigma(a)}-m_{\tilde{\sigma}(a)})
\label{complete}\ee
For definiteness, let us choose our BPS domain walls to have 
$\Delta m < 0$ where $\Delta m = \sum_{i\in \Sigma}m_i - \sum_{i\in\tilde{\Sigma}}m_i$, 
so that we pick the upper signs in this equation. The inequality above is then saturated 
if the domain walls obey the first order Bogomoln'yi equations
\be
{\cal D}_3\phi&=&e^2\left(\sum_iq_iq_i^\dagger-v^2\right) \label{bogwallone}\\
{\cal D}_3q_i&=&(\phi-m_i)q_i
\label{bogwalltwo}\ee
Configurations satisfying these equations are BPS, preserving $1/2$ of the
supersymmetries \cite{intersect,shif}. While explicit solutions to these equations
are not known in general, exact solutions can be found in the sigma-model limit $e^2\rightarrow\infty$
\cite{at,tong,sakai} and, curiously, for specific finite values of $e^2$ \cite{isozumi}.
Nevertheless, many properties of the solutions are well understood.

\para
The structure of the domain wall profile at finite $e^2$ was studied
in \cite{cobi} and \cite{shif} and is rather interesting. It depends
on the dimensionless parameter $e^2v^2/\Delta m^2$. When $e^2v^2\gg
\Delta m^2$, we are close to the sigma-model limit and we can refer
to the exact solutions: one finds that all fields interpolate
between their vacuum values over a width $1/\Delta m$. However, in
the opposite limit $e^2v^2\ll \Delta m^2$ things are somewhat different and
the domain wall exhibits a three-layer structure. In the two outer
layers, each of width $1/ev$, the fundamental fields drop to zero.
In the inner layer which has width $\Delta m/e^2v^2$, the adjoint
field $\phi$ interpolates between its two vacuum values. 
The configuration is shown in Figure \ref{fig:profile}.

\para
A heuristic understanding of how this structure arises can be gleaned
as follows \cite{shif}: the components of the quark fields $q_i$ and $q_{i+1}$
which vary over the domain wall are non-BPS with masses scaling as
$ev$. It is therefore to be expected that they vary with correlation
length $1/ev$. But when the $q_i$ vanish, equation \eqn{bogwallone}
tells us that $\phi$ changes by $\Delta m$ linearly with gradient
$e^2v^2$, resulting in the larger inner segment of the wall profile.
%%%%%%%%%%%%%%%%%%%%%%%%%%%%%%%%%%%%%%%%%%%%%%%%%%%%%%%%%%%%%%%
%%%%%%%%%%%%%%%%%%%%%%%%%%%%%%%%%%%%%
\newcommand{\onefigurenocap}[1]{\begin{figure}[h]
         \begin{center}\leavevmode\epsfbox{#1.eps}\end{center}
         \end{figure}}
\newcommand{\onefigure}[2]{\begin{figure}[htbp]
         \begin{center}\leavevmode\epsfbox{#1.eps}\end{center}
         \caption{\small #2\label{#1}}
         \end{figure}}
%%%%%%%%%%%%%%%%%%%%%%%%%%%%%%%%%%%%
\begin{figure}[htbp]
\begin{center}
\epsfxsize=3.0in\leavevmode\epsfbox{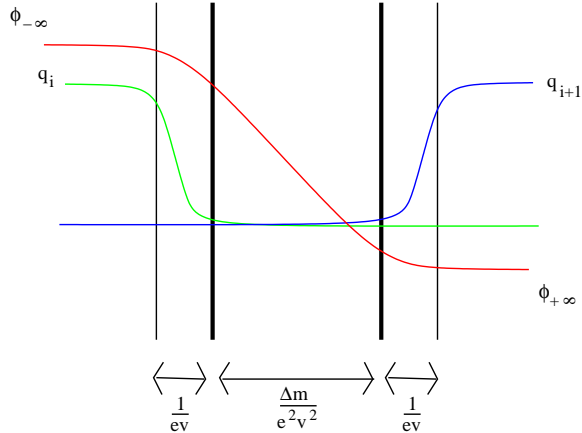}
\end{center}
 \caption{\small %Figure 1: 
The profile of an elementary domain wall 
of type $\vec{g}=\valpha_i$ when
$e^2v^2 \ll |m_i-m_{i+1}|^2$. The
fundamental fields $q$ vary in the outer layer; 
the adjoint field $\phi$ varies in 
the inner layer. Notice that, with some of the $q$'s 
vanishing, the gauge group is partially 
restored in the inner layer.}
\label{fig:profile}
\end{figure}
%%%%%%%%%%%%%%%%%%%%%%%%%%%%%%%%%%%%%%%%%%%%%%%%%%%%%%%%%%%%%%%%%%%%%%%%%%%%%%%%

\subsubsection*{\it Classification of Domain Walls}

Domain walls in field theories are classified by a choice
of vacuum $\tilde{\Sigma}$ and $\Sigma$ at left and right infinity. However,
in our theory the number of isolated vacua is $N_{\rm vac}=N_f!/N_c!(N_f-N_c)!$,
which grows exponentially with $N_f$ (fixing $N_f/N_c$), leading to a bewildering
number of possible domain walls. To help alleviate the ensuing sense of confusion,
we present here a simple
classification of the possible walls which encodes many of their properties.

\para
Firstly define the $N_f$-vector $\vec{m}=(m_1,\ldots, m_{N_f})$. From \eqn{complete},
the tension of a BPS domain wall is given by
\be
T_{\rm wall}=-v^2\,{\Tr}[\phi]^{+\infty}_{-\infty}\equiv  v^2\, \vec{m}\cdot\vec{g}\, >\, 0
\label{twall}\ee
where we have introduced the $N_f$-vector $\vec{g}\in\Lambda_R(su(N_f))$, the root
lattice\footnote{We ignore the factor
of two distinction between the root and co-root lattice of the flavour group. This
may become important in theories with non-simply laced flavour symmetries.}
of $su(N_f)$. Note that a choice of $\vec{g}$ does not imply a unique choice
of vacua $\tilde{\Sigma}$ and $\Sigma$ at left and right infinity. Nevertheless, domain
walls which share
the same $\vec{g}$ will share many of their essential properties. Note further that
there do not exist domain walls for all $\vec{g}\in \Lambda_R(su(N_f))$: the
only admissible vectors are of the form $\vec{g}=(p_1,\ldots, p_{N_f})$ with
$p_i=0$ or $\pm1$.

\para
The first question that we wish to ask is whether there exist solutions to
\eqn{bogwallone} and \eqn{bogwalltwo} for a given $\vec{g}$ and, if so,
how many parameters does the solution have. In Appendix B we perform an
index theorem calculation to determine the number of zero modes to the
domain wall equations. To express the answer, we firstly decompose
$\vec{g}$ in terms of simple roots $\vec{\alpha}_i$,
\be
\vec{g}=\sum_{i=1}^{N_f-1}n_i\vec{\alpha}_i
\nn\ee
where $n_i\in{\bf Z}.$ The basis of simple roots is determined by the
vector $\vec{m}$ which defines the positive Weyl chamber  when the
flavour group is broken to the maximal torus: $SU(N_F)\rightarrow U(1)^{N_f-1}$. Hence,
$\vec{m}\cdot\valpha_i > 0$ for all $i$.
Using our canonical basis of masses
$m_1>\ldots >m_{N_f}$, we have the simple roots
$\vec{\alpha}_1=(1,-1,0,\ldots, 0)$ and $\vec{\alpha}_2=(0,1,-1,0,\ldots,0)$ all the
way through to $\vec{\alpha}_{N_f-1}=(0,\ldots,1,-1)$. The result of Appendix B is
that solutions to the BPS domain wall equations \eqn{bogwallone} and \eqn{bogwalltwo}
exist only if $n_i>0$ for all $i$. Furthermore, the number of zero modes of a solution
is given by,
\be
\mbox{\#\ wall\ zero\ modes}=2\sum_{i=1}^{N_f-1}n_i
\label{final}\ee
This is the most general result on the counting of zero modes
in this system. It is in agreement with earlier results 
including a Morse theory calculation in the sigma-model limit 
\cite{multi}, an index theorem for the abelian theory \cite{keith} 
and an analysis of the zero modes in the non-abelian theory \cite{sakai}. 
A Morse theory calculation in a different model of domain walls can be found 
in \cite{boris}.

\para
Equation \eqn{final} has a simple physical interpretation: a domain
wall labelled by $\vec{g}$ can be thought of as constructed from
$\sum_in_i$ ``elementary'' domain walls, each labelled by a simple
root $\alpha_i$. Each of these elementary domain walls has two
collective coordinates: a position in the $x^3$ direction, and a
phase. The position is self-evident and ensures that there exist
solutions which look like $\sum_in_i$ well-separated elementary
domain walls, with the distances between them arbitrary moduli. The
phase is a little more subtle. It is required by supersymmetry (the
domain wall preserves one half of ${\cal N}=2$ supersymmetry so must
have a K\"ahler moduli space) and was first explained in \cite{at}.
Recall from \eqn{breaking} that each vacuum of our theory preserves
a $G=S[\,U(1)^{N_c}_{\rm diag}\times U(1)^{N_f-N_c}]$ symmetry.
However, the choice of which $U(1)^{N_c}\subset SU(N_f)$ gets locked
with the gauge group differs from vacuum to vacuum. When a domain
wall interpolates between two vacua, $G$ is broken by the field
configuration. 
Acting on the soliton with $G$ results in Goldstone 
modes localised on the wall. 
There are also further massless, quasi-Goldstone modes predicted by the 
index theorem which do not arise from symmetries. More details on this 
phase collective coordinate can be found in \cite{tong,shif,sakai}.

\para
The positions of the elementary domain walls are not allowed 
to be arbitrary: unlike solitons of higher co-dimensions, 
domain walls must obey at least some ordering on the line. 
In the abelian theory, this ordering is absolute and domain 
walls are not able to pass through each other \cite{multi,tong}. 
In the non-abelian theory, there is (literally) room for 
manoeuvre and certain domain walls are allowed to pass through 
each other \cite{sakai}. Such walls are said to be 
``penetrable''. 
Our simple classification of domain walls also captures this 
quality. Translating the analysis of \cite{sakai}, two 
elementary domain walls, $\vec{\alpha}_i$ and $\vec{\alpha}_j$,
consecutive in space, may pass through each other whenever 
$\vec{\alpha}_i\cdot\vec{\alpha}_j=0$. It is easy to motivate 
%that 
this result by noting that domain walls with this property 
sit in non-neighbouring parts of the flavour group and do 
not interact with each other; a linear superposition of two 
such domain walls yields a new solution. 

\para
Note that the expression \eqn{final} is reminiscent of the 
formula for the number of zero 
modes of monopoles in higher rank gauge groups \cite{erick}. 
Indeed, the relationship 
between domain walls of this type and magnetic monopoles has 
been commented on by 
several authors (see, for example, \cite{at,nick}). 
The physical reason behind this 
similarity was explained in \cite{monflux} and will be 
reviewed in Section \ref{sc:vortices}. 

\para
Finally a comment about the string theory realization of these domain walls. 
In \cite{LT} it was shown how the abelian domain walls 
appear as kinky D-strings in the D1-D5 system. More recently, the BPS wall 
solutions in the non-abelian gauge theory have been realized in a similar brane 
construction using fractional D-branes on orientifolds \cite{brncst}. 
This brane construction is very convenient to understand 
and to visualize various properties of non-abelian walls 
such as penetrable versus impenetrable pairs of walls. 

\subsubsection*{\it An Example: $U(2)$ with $N_f=4$ Flavours}

The simplest non-abelian gauge theory whose domain wall system demonstrates many of
the interesting features has $N_c=2$ colours and $N_f=4$ flavours. This was also
the example discussed by Shifman and Yung in \cite{shif2}, although they considered
the non-generic case of coincident masses so found somewhat different physics
from that described below. The theory
has $4!/2!2!=6$ vacua, giving rise to $(6\times 5)/2=15$ different domain walls. Let us
describe the properties of all of these walls in terms of our classification.

\para
For six choices of vacua at left and right infinity, there is only a single elementary
domain wall interpolating between them. These choices are:
\be
\tilde{\Sigma}=(1,2)\ \ ,\ \ \Sigma=(1,3)\ \ \ &{\Rightarrow}&\ \ \vec{g}=\vec{\alpha}_2 \nn\\
\tilde{\Sigma}=(1,3)\ \ ,\ \ \Sigma=(2,3)\ \ \ &{\Rightarrow}&\ \ \vec{g}=\vec{\alpha}_1 \nn\\
\tilde{\Sigma}=(1,3)\ \ ,\ \ \Sigma=(1,4)\ \ \ &{\Rightarrow}&\ \ \vec{g}=\vec{\alpha}_3 \nn\\
\tilde{\Sigma}=(1,4)\ \ ,\ \ \Sigma=(2,4)\ \ \ &{\Rightarrow}&\ \ \vec{g}=\vec{\alpha}_1 \nn\\
\tilde{\Sigma}=(2,3)\ \ ,\ \ \Sigma=(2,4)\ \ \ &{\Rightarrow}&\ \ \vec{g}=\vec{\alpha}_3 \nn\\
\tilde{\Sigma}=(2,4)\ \ ,\ \ \Sigma=(3,4)\ \ \ &{\Rightarrow}&\ \ \vec{g}=\vec{\alpha}_2
\nn\ee
Each of these domain wall systems has just two collective coordinates: a center of mass
and a Goldstone phase.

\para
For five choices of vacua at left and right infinity, the domain wall system decomposes
into two elementary domain walls. These choices are:
\be
\tilde{\Sigma}=(1,2)\ \ ,\ \ \Sigma=(1,4)\ \ \ &{\Rightarrow}&\ \ \vec{g}=\vec{\alpha}_2+\vec{\alpha}_3 \nn\\
\tilde{\Sigma}=(1,2)\ \ ,\ \ \Sigma=(2,3)\ \ \ &{\Rightarrow}&\ \ \vec{g}=\vec{\alpha}_1+\vec{\alpha}_2 \nn\\
\tilde{\Sigma}=(1,3)\ \ ,\ \ \Sigma=(2,4)\ \ \ &{\Rightarrow}&\ \ \vec{g}=\vec{\alpha}_1+\vec{\alpha}_3 \nn\\
\tilde{\Sigma}=(1,4)\ \ ,\ \ \Sigma=(3,4)\ \ \ &{\Rightarrow}&\ \ \vec{g}=\vec{\alpha}_1+\vec{\alpha}_2 \nn\\
\tilde{\Sigma}=(2,3)\ \ ,\ \ \Sigma=(3,4)\ \ \ &{\Rightarrow}&\ \ \vec{g}=\vec{\alpha}_2+\vec{\alpha}_3
\nn\ee
Each of these solutions has a regime where it looks like two well-separated domain walls. For four of
the choices of vacua, the two domain walls cannot pass through each other: as they come together,
they merge to form a single domain wall. The moduli space is smooth at this point \cite{tong}. However,
the remaining choice of vacua is special: for $\vec{g}=\vec{\alpha}_1+\vec{\alpha}_3$, we have
$\vec{\alpha}_1\cdot\vec{\alpha}_3=0$ and the domain
walls are penetrable and may move through each other \cite{sakai}.

\para
There is one further choice of vacua for which the domain wall system looks like two
elementary domain walls. This is
\be
\tilde{\Sigma}=(1,4)\ \ ,\ \ \Sigma=(2,3)\ \ \ &{\Rightarrow}&\ \ \vec{g}=\vec{\alpha}_1-\vec{\alpha}_3
\nn\ee
Note the minus sign! From the previous discussion, we learn that this system is to be thought
of as a domain wall combined with an anti-domain wall. It breaks supersymmetry and there is
no solution to the first order equations \eqn{bogwallone} and \eqn{bogwalltwo} with this choice
of boundary conditions. This example of supersymmetry breaking was also noticed in \cite{sakai}.

\para
There are two choices of vacua which have a system of three domain walls interpolating between
them. They are
\be
\tilde{\Sigma}=(1,2)\ \ ,\ \ \Sigma=(2,4)\ \ \ &{\Rightarrow}&\ \ \vec{g}=\vec{\alpha}_1
+\vec{\alpha}_2+\vec{\alpha}_3 \nn\\
\tilde{\Sigma}=(1,3)\ \ ,\ \ \Sigma=(3,4)\ \ \ &{\Rightarrow}&\ \ \vec{g}=\vec{\alpha}_1+\vec{\alpha}_2+
\vec{\alpha}_3
\nn\ee
In each of these systems, one domain wall is fixed in the ordering,
while the other two may pass through
each other. For the first set of vacua, moving in the direction of positive
$x^3$, the ordering of domain walls is either
$\vec{\alpha}_2\rightarrow\vec{\alpha}_1\rightarrow\vec{\alpha}_3$ or
$\vec{\alpha}_2\rightarrow\vec{\alpha}_3\rightarrow\vec{\alpha}_1$. For
the second system, the domain wall ordering can be either
$\vec{\alpha}_1\rightarrow\vec{\alpha}_3\rightarrow\vec{\alpha}_2$
or $\vec{\alpha}_3\rightarrow\vec{\alpha}_1\rightarrow
\vec{\alpha}_2$. Notice that this example reveals that the classification
of domain wall systems in terms of $\vec{g}$ is {\it not} sufficient to
determine the ordering of the walls: different systems with the same $\vec{g}$
may have different orderings.

\para
%%%%%%%%%%%%%%%%%%%%%%%%%%%%%%%%%%%%
\begin{figure}[htbp]
\begin{center}
\epsfxsize=5.5in\leavevmode\epsfbox{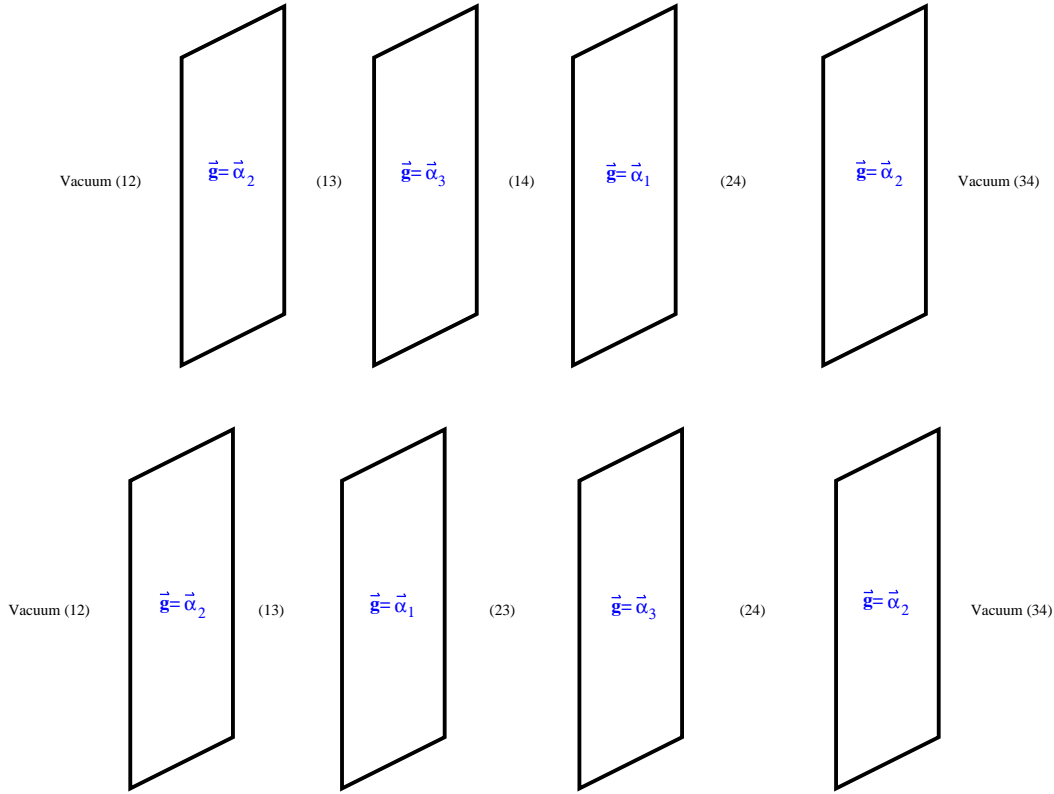}
\end{center}
 \caption{\small %Figure 2: 
The two different orderings of the domain 
wall system $\vec{g}=\vec{\alpha}_1+2\vec{\alpha}_2+\vec{\alpha}_3$. 
The two orderings are related by interchanging the inner two 
domain walls, an operation which is allowed since 
$\valpha_1\cdot\valpha_3=0$. Notice that the vacuum in the 
middle of the system changes as the walls pass through each 
other.}
\label{fig:wall-orderings}
\end{figure}
%%%%%%%%%%%%%%%%%%%%%%%%%%%%%%%%%%%%%%%%%%%%%%%%%%%%%%%%%%%%
Finally, the richest domain wall system occurs for a unique choice of vacua
at left and right infinity:
\be
\tilde{\Sigma}=(1,2)\ \ ,\ \ \Sigma=(3,4)\ \ \ &{\Rightarrow}&\ \ \vec{g}=\vec{\alpha}_1+
2\vec{\alpha}_2+\vec{\alpha}_3
\nn\ee
This system has 8 collective coordinates and may be decomposed 
into four elementary domain walls. The two $\valpha_2$ domain 
walls are constrained to lie on the outside of the system, 
while the $\valpha_1$ and $\valpha_3$ domain walls are free 
to exchange their positions in the middle. 
The two different orderings of the domain wall solutions 
are drawn in Figure \ref{fig:wall-orderings}.

\subsubsection*{\it The Low-Energy Dynamics of Walls}

As we saw in equation \eqn{final}, a domain wall system, classified
by $\vec{g}$, has ${\cal I}=\sum_in_i$ complex collective
coordinates which we shall denote as $X^m$, $m=1,\ldots {\cal I}$.
The $X^m$ are coordinates on the domain wall moduli space ${\cal
M}_{\rm wall}$. The low-energy dynamics of the domain walls is
derived by promoting these collective coordinates to dynamical
fields on the domain wall worldvolume, $X^m\rightarrow
X^m(t,x^1,x^2)$, and plugging the resulting field content back into
the Lagrangian \eqn{lag}. Performing the integral over the $x^3$
direction transverse to the domain wall results in a $d=2+1$
dimensional sigma-model with target space ${\cal M}_{\rm wall}$. The
metric on ${\cal M}_{\rm wall}$ is determined by the kinetic terms
of the original Lagrangian \cite{manton} and, thanks to
supersymmetry, is K\"ahler.

\para
Consider a single elementary domain wall $\vec{g}=\valpha_i$. 
The tension of such a wall is given by 
$T_{\rm wall}=v^2(m_i-m_{i+1})$ and the solution has a 
single complex collective coordinate: 
\be X = x^3_{(0)}+i\theta \label{dothedual}\ee
where $x^3_{(0)}$ is the center of mass of the domain wall 
in the $x^3$ direction, and $\theta$ is the phase of the 
wall arising from acting with the symmetry group 
\eqn{breaking} preserved in the vacuum. 
It has periodicity $\theta\equiv\theta+2\pi v^2/T_{\rm wall}$. 
The low-energy dynamics of a single elementary wall is 
simply described by the following effective Lagrangian on 
the wall worldvolume 
\be 
-{\cal L}_{\rm wall} =  \ft12 
T_{\rm wall} \partial_\mu \bar X \partial^\mu X, 
\label{firstx}
\ee
where $\mu=0,1,2$ runs over the worldvolume of the domain 
wall. 
%$z=x^1+ix^2$ and $\partial = \partial/\partial z$. 
There is a dual formalism of the domain wall dynamics 
noticed in \cite{dbrane}. (See \cite{shif} for a careful 
discussion in the case with finite coupling $e^2$). 
In $d=2+1$ dimensions, a periodic scalar field is dual to 
a $U(1)$ photon\footnote{We use the same 
letter $F$ to denote the $U(N_c)$ bulk field and the $U(1)$ 
worldvolume field and hope the context will suffice to 
alleviate any confusion.}:
\be 
4\pi %\frac
{v^2}%{2T_{\rm wall}}
\partial_\mu \theta = 
\epsilon_{\mu\nu\rho}F^{\nu\rho} 
\label{donthedual}
\ee
where 
the normalisation is chosen so that a vortex in the $\theta$ 
field corresponds to a unit electric charge in the field 
strength $F$. 
We can therefore re-write the theory on the domain wall in 
terms of a single neutral scalar field $x^3_{(0)}$ and a 
$U(1)$ gauge field $F_{\mu\nu}$ 
\be 
-{\cal L}_{\rm wall} = \ft12 T_{\rm wall} 
\partial_\mu x^3_{(0)} \partial^\mu x^3_{(0)} 
+
\frac{(m_i-m_{i+1})}{16%2
v^2\pi^2}F_{\mu\nu}F^{\mu\nu}. 
\label{dual}
\ee
As explained in \cite{shif}, in the regime of parameters 
$e^2v^2\ll m^2$, the gauge field on the domain wall can also 
be understood in terms of the Dvali-Shifman localisation 
mechanism \cite{dvali} since, as we see from 
Figure \ref{fig:profile}, 
the profile of the domain wall 
results in an unbroken gauge group in the center. The fact that the
gauge group is Higgsed in the bulk (as opposed to confined) means
that the dual bulk gauge field is localised on the wall
\cite{schmaltz}. As a check of this interpretation, note that the
effective $d=2+1$ gauge coupling constant 
$e^2_{2+1}$ in \eqn{dual} scales as
\be
\frac{1}{e^2_{2+1}}\sim\frac{(m_i-m_{i+1})}{v^2}
\sim e^2\times(\mbox{width of domain wall})
\nn\ee
\para
The domain wall also has fermionic zero modes. 
These complete the above Lagrangian to one 
with ${\cal N}=2$ supersymmetry in $d=2+1$ dimensions 
(that is, four supercharges). Thus 
the low-energy effective action for a single domain wall 
is a free ${\cal N}=2$ $U(1)$ 
gauge theory. In Section \ref{sc:d-branes}%6
, we shall see how charged particles 
enter this picture.

\para
For multiple domain walls, expressions for the metric on the 
moduli space ${\cal M}_{\rm wall}$ in abelian theories 
were computed in \cite{tong} (see also \cite{mirror,isozumi}).
For example, for two elementary domain walls 
$\vec{g}=\valpha_i+\valpha_{i+1}$, the moduli 
space is of the form 
${\cal M}_{\rm wall}\cong {\bf R}\times 
({\bf S}^1\times \tilde{\cal M})/{\bf Z_2}$
where $\tilde{\cal M}$ is a manifold with the topology of 
a cigar, corresponding to the relative position and phase 
of the two domain walls. For domain walls in non-abelian 
theories, a description of the domain wall moduli space was 
presented in \cite{sakai} in terms of the underlying 
Grassmannian $G(N_c,N_f)$ with subspaces removed 
corresponding to ``points at infinity''.

\section{Vortices}
\label{sc:vortices}

Our theory also admits another type of classical soliton 
solution: a vortex string. In the abelian 
case, these are simply the Abrikosov-Nielsen-Olesen vortices 
\cite{no}. (For $N_f>1$ flavours, they are 
known as semi-local vortices \cite{vash}). 
In non-abelian gauge theories, the dynamics of 
vortices was studied in detail in \cite{vib,yung}. 
This section reviews the main results.

\para
The Bogomoln'yi equations for vortices can be derived by 
completing the Hamiltonian \eqn{ham} in a different manner 
from the previous section. We will consider vortex strings 
lying in the $x^3$ direction, with the profile in the 
$x^1, x^2$ plane. 
So we restrict to configurations with 
$\partial_3=A_3=0$. The string carries magnetic flux $\int dx^1dx^2{\rm Tr}B_3=-2\pi k$ 
where $k\in {\bf Z}$. We have
\be 
\int dx^1dx^2\ {\cal H}&=&\int dx^1dx^2\ 
 \left[ \frac{\Tr}{2e^2}(B_3\mp e^2
(\sum_{i=1}^{N_f}q_iq_i^\dagger-v^2))^2
+\frac{1}{2e^2}\,{\rm Tr}\left[({\cal D}_1\phi)^2 + ({\cal D}_2\phi)^2\right]\right. 
\nn\\
&& \ \ \ \ \ \ 
\left. + \sum_{i=1}^{N_f}q_i^\dagger(\phi-m_i)^2q_i+
\sum_{i=1}^{N_f}\left|({\cal D}_1\mp i{\cal D}_2) q_i\right|^2 
\mp v^2\,\Tr\,B_3\right]
\nn\\ &\geq & \mp v^2\,\int dx^1dx^2\ \Tr\, B_3 
=\pm 2\pi v^2 k, 
\nn\ee
where we choose the upper (lower) sign for $k>0$ ($k<0$). For the remainder of this 
section we restrict to $k>0$, but the possibility for both signs will become important when 
we consider the interaction of vortex strings with domain walls. The inequality above is 
saturated for BPS vortices of tension $T_{\rm vortex}=2\pi v^2|k|$ which satisfy the 
non-abelian first-order vortex equations,

\be
B_3=e^2(\sum_{i=1}^{N_f}q_iq_i^\dagger -v^2)\ \ \ \ 
&;&\ \ \ \ \ {\cal D}_z q_i=0 
\label{vortone}\\
(\phi-m_i)q_i=0 \ \ \ \ \ &;& \ \ \  \ 
{\cal D}_1\phi= {\cal D}_2 \phi=0. 
\label{vorttwo}
\ee
where $z=x^1+ix^2$ labels the plane transverse to the vortex string. 
The space of solutions to these equations is richest 
when the masses vanish: $m_i=0$. 
We will describe this situation first, then move on to explain 
how it is changed when the masses are turned on. 
When $m_i=0$, the second line of equations \eqn{vorttwo} is 
trivially solved by $\phi=0$ and we can focus on the first line. 
The number of zero modes of 
these equations was computed in \cite{vib}, with the result 
%%%%%%%%%%%%%%%%%%%%%%%%%%%%%%%%%%%%
\begin{figure}[htbp]
\begin{center}
\epsfxsize=4.0in\leavevmode\epsfbox{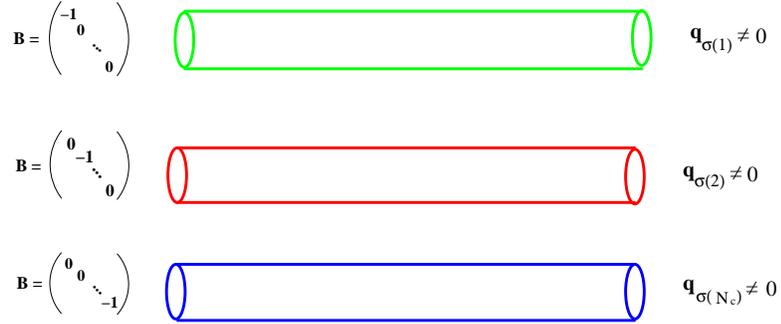}
\end{center}
 \caption{\small %Figure 3: 
The $N_c$ different types of vortices, 
labelled by the element of the magnetic field or, equivalently, 
by the flavour $q_i$ which winds around the vortex. 
The $N_c$ different vortex strings 
are mutually BPS.}
\label{fig:vortices}
\end{figure}
%%%%%%%%%%%%%%%%%%%%%%%%%%%%%%%%%%%%%%%%%%%%%%%%%%%%%%%%%
\noindent
\be
\mbox{\#\ vortex\ zero\ modes}=2 k N_f
\label{vortexcount}\ee
which generalises the answer in the abelian case computed in 
\cite{erickvort}. 
As with most BPS solitons, the proportionality to $k$ can be 
understood as reflecting the fact that there exist solutions 
corresponding to $k$ well-separated charge $1$ 
vortices. Each of these has $2N_f$ zero modes which decompose 
as follows: two of these zero modes are simply the position 
of the vortex on the $(x^1,x^2)$ plane. 
Of the remaining collective coordinates, some are Goldstone 
modes. To see this, note that when the masses vanish $m_i=0$, 
the symmetry breaking pattern \eqn{breaking} 
is replaced by
\be
U(N_c)\times SU(N_f)
\rightarrow S[U(N_c)_{\rm diag}\times U(N_f-N_c)]
\label{newsym}\ee
A single vortex generically breaks this surviving symmetry 
group, meaning 
that we can sweep out new vortex solutions by acting with it. 
Finally, there 
are also collective coordinates corresponding to approximate 
scaling modes (see for example \cite{nonnorm}). 
We shall discuss the vortex moduli space in more 
detail presently.

\para
How does this situation change when the masses $m_i\neq 0$? 
The answer to this question was given in \cite{monflux} 
(see also \cite{misha,vstring}). 
Given that some of the vortex zero modes are 
generated by acting with the symmetry \eqn{newsym} which is no
longer available when masses are turned on \eqn{breaking}, 
it should be immediately clear that much of the moduli 
space is lifted. In fact, the only surviving solutions to 
\eqn{vortone} and \eqn{vorttwo} are abelian vortex solutions 
embedded diagonally within the gauge group,
\be (B_3)^a_{\ b}
=B^a_\star\delta^a_{\ b} \ \ \ \ ,\ \ \ \ \ q^a_{\
i}= q^a_\star \delta^a_{\ i=\sigma(a)} \label{bstar}
\ee
(no sum over $a$), where each pair $\{B_\star^a,q_\star^a\}$ 
for $a=1,\ldots,N_c$ solves the abelian vortex equations. 
In this manner, a given vacuum contains $N_c$ different 
types of vortex string, labelled by the diagonal element of 
the gauge group which carries the magnetic flux or, 
relatedly, by the flavour of quark 
field which has non-zero winding at infinity. 
The single topological quantum number $k$ splits into $N_c$ distinct quantum numbers 
$k_a\in{\bf Z}>0$, where $\sum_ak_a=k$. (The gauge invariant 
object is $\int \Tr\,\phi B$). 
The vortices in each of these sectors have 
$2k_a$ zero modes, corresponding to the positions of 
$k_a$ vortex strings in the $(x^1,x^2)$ plane. 
All internal zero modes have been 
lifted by the masses. The $N_c$ different strings are drawn 
for a given vacuum in Figure \ref{fig:vortices}%3
. 
We shall refer to the string around which 
the phase of $q_i$ winds as the ``$q_i$-string''. 
It only exists in a given vacuum if $i\in\Sigma$.

\subsubsection*{\it The Low-Energy Dynamics of Vortices}

When the masses vanish ($m_i=0$),
the low-energy dynamics of vortex strings is derived
by promoting the $2 k N_f$ collective coordinates to 
dynamical fields on the string worldsheet 
(i.e. depending on $t$ and $x^3$). Plugging the 
resulting configuration into the Lagrangian \eqn{lag} 
and performing the integral over the $(x^1,x^2)$ plane 
results in a $d=1+1$ dimensional sigma-model on the vortex 
moduli space ${\cal M}_{\rm vortex}$ where the 
result \eqn{vortexcount} tells us 
${\rm dim}({\cal M}_{\rm vortex})=2kN_f$. 
Since the vortex is $1/2$-BPS, the $d=1+1$ vortex theory 
has ${\cal N}=(2,2)$ supersymmetry (four supercharges); 
the superpartners of the bosonic zero modes are the 
(non-chiral) fermionic zero modes of the vortex string.

\para
An example: for a single $k=1$ vortex in $U(N_c)$ gauge 
theory with $N_f=N_c$ flavours, the vortex moduli space 
is \cite{vib,yung} 
\be
{\cal M}_{\rm vortex} \cong {\bf C}\times {\bf CP}^{N_c-1}
\label{cpn}
\ee
where the K\"ahler class (or size) of ${\bf CP}^{N_c-1}$ is given by $r=2\pi/e^2$. 
For higher $k$ and higher $N_f$, the topology of the vortex 
moduli space was determined in \cite{vib}.

\para
When the masses are non-zero, we have seen that much of the 
vortex moduli space is lifted. This can be understood as 
introducing a potential on ${\cal M}_{\rm vortex}$ 
\cite{monflux,misha,vstring}. 
Here we present a description of the resulting dynamics on 
the vortex string worldsheet in terms of a $d=1+1$ 
${\cal N}=(2,2)$ gauged linear sigma model, derived using 
D-brane techniques in \cite{vib}. 
Consider a single $k=1$ vortex string sitting in a given 
vacuum $\Sigma$. The worldvolume dynamics is described by 
a $U(1)$ vector multiplet containing a gauge field $G_{03}$ 
and a neutral complex scalar $\chi$. 
Coupled to this are $N_c$ chiral multiplets
$\psi_a$ of charge $+1$, and a further $(N_f-N_c)$ chiral 
multiplets $\hat{\psi}_m$ of charge $-1$. 
These charged chiral multiplets 
parameterise the internal degrees of freedom of the vortex. 
Finally, there is a single neutral chiral multiplet $z$ 
containing the center of mass of the vortex. 
Denoting the lowest components of the chiral 
superfields by the same letters $\psi,\hat{\psi}$ and $z$, 
the bosonic part of the Lagrangian describing the dynamics 
of a single vortex is:
\be 
-{\cal L}_{\rm vortex} &=&
\frac{1}{2g^2}\left(G_{03}^2+|\partial\chi|^2\right)
+\sum_{a=1}^{N_c}\left(|{\cal D}\psi_a|^2 +
|\chi-m_{\sigma(a)}|^2|\psi_a|^2
\right) + |\partial z|^2 \label{ours}\\
&+& \sum_{m=1}^{N_f-N_c}\left(|{\cal D}\hat{\psi}_m|^2+
|\chi-m_{\hat{\sigma}(m)}|^2|\hat{\psi}_m|^2\right)
+\frac{g^2}{2}(\sum_{a=1}^{N_c}|\psi_a|^2
-\sum_{m=1}^{N_f-N_c}|\hat{\psi}_m|^2-r)^2, 
\nn
\ee
where $\hat{\Sigma}$ is a set of $(N_f-N_c)$ elements 
$\hat{\sigma}(m)$, defined to be the complement of $\Sigma$ 
in $\{1,\ldots,N_f\}$, and 
$|\partial\chi|^2\equiv -(\partial_0\chi)^2+(\partial_3\chi)^2$. 
This Lagrangian contains three parameters: the 
masses $m_{\sigma(a)}$ and $m_{\hat{\sigma}(m)}$ are set 
equal to the masses $m_i$ of the four-dimensional theory 
(strictly speaking,
these are referred to as ``twisted masses'' in two dimensions 
\cite{twisted}). 
The FI parameter $r$ on the vortex worldsheet is
given by \cite{vib}
\be r=\frac{2\pi}{e^2} \label{r}\ee
Finally the gauge coupling is to be taken to infinity 
$g^2\rightarrow\infty$, rendering the vector multiplet 
auxiliary since its kinetic term vanishes. 
As reviewed in Appendix A, this ensures that attention 
is restricted to the Higgs branch of the 
theory given by the vanishing of the D-term 
($\sum_a|\psi_a|^2-\sum_m|\hat{\psi}_m|^2=r)$, 
modulo the $U(1)$ gauge action. 
This Higgs branch is the vortex moduli space. 
The masses $m_i$ induce a potential on this Higgs branch 
as anticipated in the above discussion.

\para
An example: When $N_f=N_c$ we have no $\hat{\psi}$ fields. 
The D-term is simply $\sum_a|\psi_a|^2=r$ which, after 
quotienting by the $U(1)$ action 
$\psi_a\rightarrow \exp(i\alpha)\psi_a$ results in 
the Higgs branch ${\bf CP}^{N_c-1}$. If the masses are 
zero $m_i=0$ then this Higgs branch is not lifted and, 
including the neutral field $z$, the vacuum moduli space of 
the theory \eqn{ours} is given by 
${\cal M}_{\rm Higgs}\cong {\bf C}\times {\bf CP}^{N_c-1}$ 
in agreement with the vortex moduli space \eqn{cpn}.

\para
When the twisted masses are turned on, the vortex theory 
\eqn{ours} has $N_c$ discrete vacua given by,
\be
{\rm Vacuum}\ a: \ \ \ \chi=m_{\sigma(a)}\ \ \ \ ,\ \ \ \
|\psi_b|^2=r\delta_{ab}\ \ \ \ , \ \ \  \ \ \hat{\psi}_m = 0
\label{vac}\ee
This reflects the fact that, as explained above, the $k=1$ 
vortex has $N_c$ ground states determined by which scalar 
field $q$ winds at infinity. 
The vacuum $a$ of the vortex theory describes the situation 
in which the $q_{i=\sigma(a)}$ scalar field carries the 
winding and the magnetic field sits in the $a^{\rm th}$ 
diagonal component of $B_3$. 

\para
For $k$ vortices, the low-energy dynamics is described by a 
$U(k)$ gauge theory with an adjoint chiral multiplet, 
$N_c$ fundamental chiral multiplets and $(N_f-N_c)$ 
anti-fundamental chiral multiplets. 
The derivation of this theory, together with a discussion of 
the physics it captures, can be found in \cite{vib}.

\section{Confined Monopoles}
\label{sc:confined-monopoles}

So far we have discussed the domain wall and vortex string 
solutions of the theory. 
Now it is time to move onto monopoles. From the
symmetry breaking pattern \eqn{breaking}, there is no unbroken
$U(1)$ of the gauge symmetry and, correspondingly, no massless
photon which can carry away the flux of a 't Hooft-Polyakov
monopole. Nevertheless, as shown in \cite{monflux}, BPS 
monopoles do exist in this theory. 
Since they lie in the Higgs phase, their flux
is carried away in collimated tubes. 
In other words, they are 
confined. In this section we review the results of 
\cite{monflux}.
The quantisation of these objects was considered in 
\cite{misha,vstring} while discussions of confined monopoles 
in other theories can be found in \cite{other}.

\para
The confined monopole is a $1/4$-BPS object. 
The first order Bogomoln'yi equations can be once again 
derived from completing the square in the Hamiltonian 
\eqn{ham}. This time we have two choices of $\pm$ signs. To make 
this clear, define $\epsilon$ and $\eta$ to independently\footnote{
The sign $\eta$ ($\epsilon\eta$) corresponds to the sign choice 
%of the gamma matrix $\gamma_{\rm wall}$ ($\gamma_{\rm monopole}$) 
for the wall (the vortex) when Bogomoln'yi completion is performed. 
%projection operator. 
} take 
values $\pm 1$. Then we can write 
\be 
\int d^3x\,{\cal H}&=&\int d^3x\, \frac{1}{2e^2}\Tr\Biggl[
({\cal D}_1\phi +\epsilon B_1)^2 
+ ({\cal D}_2\phi+\epsilon B_2)^2 %\right.
\nn \\
&& \ \ \ \ \ \ 
%\left.
+({\cal D}_3\phi+\epsilon B_3 
-\eta e^2(\sum_iq_iq_i^\dagger-v^2))^2\Biggr] 
\nn\\ 
&& \ \ \ \ \ \ 
+\sum_i\left|({\cal D}_1 -\epsilon \eta i{\cal D}_2)q_i\right|^2 
+\sum_i\left|{\cal D}_3q_i - \eta (\phi-m_i)q_i\right|^2 
\nn\\ 
&& \ \ \ \ \ \ +  \Tr\left[ -\eta v^2\partial_3\phi -\epsilon\eta v^2B_3 
-\frac{\epsilon}{e^2}\partial_\alpha(\phi B_\alpha)\right] \nn\\ 
&\geq&
\int\,d^3x\ \Tr\,\left[\,-\eta v^2\partial_3\phi -\epsilon\eta v^2B_3 
-\frac{\epsilon}{e^2}\partial_\alpha(\phi B_\alpha)\right] \nn\\ 
&\equiv&
\left[\int dx^1dx^2\ T_{\rm wall}\right] 
+ \left[\int dx^3\ T_{\rm vortex}\right] + M_{\rm mono}
\label{finalbog}
\ee
%
%%%%%%%%%%%%%%%%%%%%%%%%%%%%%%%%%%%%
\begin{figure}[htbp]
\begin{center}
\epsfxsize=5.5in\leavevmode\epsfbox{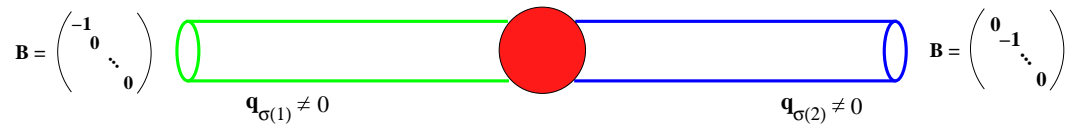}
\end{center}
 \caption{\small %Figure 4: 
The confined monopole which is a source 
for magnetic flux 
$\int d^2x B=2\pi \ {\rm diag}(1,-1,\ldots,0)$.}
\label{fig:monflux}
\end{figure}
%%%%%%%%%%%%%%%%%%%%%%%%%%%%%%%%%%%%%%%%%%%%%%%%%%%%%%%%%%%%%
In the final line the central charges for 
domain walls and vortices appear 
which are familiar from Sections \ref{sc:domain-wall} %3 
and 4 respectively. The monopole central charge also makes an 
appearance,
\be
M_{\rm mono}=-\frac{\epsilon}{e^2}\,{\rm Tr}\int\partial_\alpha(\phi B_\alpha)
\label{mono}\ee
In this section, we shall set 
$T_{\rm wall}=0$ by working in the same vacuum $\Sigma$ for 
all spatial infinity. In the following sections we shall relax this 
restriction to examine the most general configurations 
which involve monopoles, vortices and walls. 

\para
If we follow the convention of Section 3 and 4, we set $\eta=\epsilon=+1$, have the 
Bogomoln'yi equations
\be 
 B_1=-{\cal D}_1\phi
\ \ \ ,\ \ \ B_2=-{\cal D}_2\phi=0\ \ \ ,\ \ \
B_3=-{\cal D}_3\phi+e^2(\sum_{i=1}^Nq_iq_i^\dagger -v^2)
\label{boggy1}
\\
{\cal D}_1q_i=i{\cal D}_2 q_i\ \ \ ,\ \ \ 
{\cal D}_3 q_i=(\phi-m_i)q_i. \hspace{2.5cm} 
\label{boggy}
\ee
Other choices of minus sign are self-evident.
The equations \eqn{boggy1} and \eqn{boggy} are 
over-constrained: $2N_c(2N_c+N_f)$ real degrees
of freedom, related by $N_c^2$ gauge transformations and
$N_c(3N_c+4N_f)$ real differential equations. Consistency of 
the equations follows by noting the integrability condition \cite{allquarter2}  
\be 
\left[{\cal D}_1 - i{\cal D}_2, {\cal D}_3- \phi+m_i\right] 
=-\left(B_1 + {\cal D}_1\phi\right) 
+ i \left(B_2 + {\cal D}_2\phi\right)
=0. 
\label{eq:integrability}
\ee
This ensures that the two equations in \eqn{boggy} can be solved 
simultaneously for $q_i$. A similar integrability condition exists for 
1/4-BPS equations describing instantons in the Higgs phase \cite{EINOS}.

\para
No explicit solutions to \eqn{boggy1} and \eqn{boggy} are known. By examining the 
configuration from several perspectives, it was argued in \cite{monflux} that solutions 
for equations \eqn{boggy1} and \eqn{boggy} exist 
and several properties were derived. In \cite{misha} a radial ansatz 
was employed to find a solution perturbatively in 
$\Delta m^2/e^2v^2$. 
The solutions describe a magnetic monopole 
emitting two vortex flux tubes as sketched in 
Figure \ref{fig:monflux}. 
The energy of this entire configuration is infinite due to the presence of the two
semi-infinite vortex strings. This is the statement that the
monopole is confined. However, it still makes sense to talk about
the finite mass $M_{\rm mono}$ of the monopole as an excitation over
the energy of a single infinite, straight string. (This point will
be made clearer shortly when we examine the monopole from the
perspective of the vortex string). Recall that in the Coulomb phase
the magnetic flux of the monopole escapes radially to infinity and
is captured by the integral $M_{\rm mono}=-(1/e^2)\Tr\,\int d^3x \
\partial\cdot(\phi B)$ evaluated on  the $S^2_\infty$ boundary. In
the present case, the magnetic flux does not make it to all points
on the boundary, but is confined to two flux tubes which stretch in
the $\pm x^3$ direction. Correspondingly, the integral should now be
evaluated over two planes ${\bf R}^2_{\pm\infty}$ at
$x^3=\pm\infty$. Nevertheless, both integrals yield the same result.

\para
The choice of minus signs $\epsilon$ and $\eta$ when completing the square in 
\eqn{finalbog} determine the direction in which the vortex strings are emitted 
from the monopole. For example, with $\epsilon=\eta=+1$ we have BPS 
vortices with negative flux $\int {\rm Tr} B_3<0$. The requirement that the 
monopole mass $M_{\rm mono}=-\left[\Tr\,(\phi\,\int B_3)\right]^{\infty}_{-\infty}$ 
is positive then fixes the direction in which the vortex strings lie.  Since 
the masses are ordered as $m_i>m_{i+1}$, we learn that the confined monopole satisfying 
(\ref{boggy1}) and \eqn{boggy} must have a $q_{k}$ 
vortex string emitted to the left and a $q_j$ 
string emitted to the right with $k>j$.  In terms of 
Figure \ref{fig:monflux}, we require $\sigma(1)>\sigma(2)$. 
Other choices of $\epsilon$ and $\eta$ 
allow for a reversal of the string orientation but, crucially, 
a choice of $\eta$ is required for compatibility for domain walls 
as we shall see in Section \ref{sc:interaction}.

\para
Let us remind ourselves how to classify magnetic monopoles in higher
rank gauge groups. The magnetic charge lives in the Cartan
subalgebra $U(1)^{N_c-1}\subset SU(N_c)$ of the gauge group and is
defined by a charge vector $\vec{h}\in \Lambda_R(su(N_c))$, the
(co)root lattice\footnote{Again, we ignore the factor of two
difference between roots and co-roots which is unimportant for simply
laced groups.} of the Lie algebra \cite{gno}. The vector $\vec{h}$
is $N_c$-dimensional, with integer components whose sum vanishes,
imposing the requirement that the magnetic charge lives in
$SU(N_c)\subset U(N_c)$. As is usual for BPS states, the mass of the
monopole is determined by the topological charge $\vec{h}$.
Recalling the vacuum expectation value $\phi={\rm
diag}(m_{\sigma(1)},\ldots,m_{\sigma(N_c)})$, we define the $N_c$
dimensional vector
$\vec{\phi}=(m_{\sigma(1)},\ldots,m_{\sigma(N_c)})$. Then the mass
of the monopole is
\be M_{\rm mono} = \frac{2\pi}{e^2}\,
\vec{h}\cdot\vec{\phi} \label{mmono}\ee
As we mentioned above, this
is the monopole mass in both the Coulomb and Higgs phases. Of
course, in the latter phase this finite energy is supplemented by an
infinite contribution from the vortex string.

\para
Finally, we recall the result of E. Weinberg on counting 
the number of zero modes of the monopole \cite{erick}. 
If we decompose the magnetic charge into simple roots 
$\vbeta_a$, with $a=1,\ldots,N_c-1$ such that 
$\vec{h}=\sum_an_a\vbeta_a$, then BPS 
monopoles exist for $n_a\in{\bf Z}>0$. 
The number of zero modes of a solution {\it in the Coulomb phase} 
is $4\sum_an_a$ \cite{erick}. 
As we commented in Section \ref{sc:theory}%2
, there are strong parallels between the classification and 
the counting of zero modes of monopoles and domain walls. 
This coincidence finds a compelling physical explanation 
when viewed from the perspective of the vortex string 
 as we now review.

\subsubsection*{\it The View from the String}

It is interesting to ask how the confined monopole looks from the
worldsheet of the vortex string. In the previous section we saw that
the low-energy dynamics of a single string is described by a $U(1)$
gauge theory with certain matter content. The Lagrangian was given
in equation  \eqn{ours}. This theory has $N_c$ vacuum states
\eqn{vac} corresponding to the $N_c$ different vortex strings that
exist in the theory (see Figure \ref{fig:vortices}). 
In summary, we have $d=1+1$
dimensional theory on the string worldsheet with a set of isolated,
discrete vacuum states. This immediately guarantees us that we have
a new object: a kink on the string. This kink is the confined
monopole \cite{monflux}.

\para
In fact, the kink is of precisely the type discussed in 
Section \ref{sc:theory}%2
; the theory in
question lives in $d=1+1$ dimensions rather than $d=3+1$, and has four supercharges
rather than eight, but the Bogomoln'yi equations for the kink on the string
have the same form as those describing the domain walls in 
Section \ref{sc:theory}%2
, namely
\be
\partial_3\chi=-g^2\left(\sum_{a=1}^{N_c}|\psi_a|^2-r\right)
\ \ \ \ \ ,\ \ \ \ \ \
{\cal D}_3\psi_a=-(\chi-m_{\sigma(a)})\psi_a
\label{kinkonstring}
\ee
supplemented with $\hat{\psi}=0$. 
These are to be compared with \eqn{bogwallone} and 
\eqn{bogwalltwo}. 
Note that the $SU(N_c)_{\rm diag}$ group in spacetime has 
descended to the flavour group on the string worldsheet. 
The extra minus signs in \eqn{kinkonstring} can be traced 
to the fact that the orientation of the emitted 
strings discussed above: the vortex theory is in the vacuum 
$\chi=m_k$ as $x_3\rightarrow -\infty$ and $\chi=m_j$ as 
$x^3\rightarrow +\infty$ with $k>j$ which implies that $m_k<m_j$.

\para
From Section \ref{sc:domain-wall}%3
, recall that we can classify kinks on the string in
terms of a root vector of the flavour group which, in this case, is
$\vec{h}\in \Lambda_R(su(N_c))$ where
$[\chi]^{+\infty}_{-\infty}=\vec{\phi}\cdot\vec{h}$. The mass of the
kink is then given by (c.f. equation \eqn{twall})
\be M_{\rm kink} = r\,
[\chi]^{+\infty}_{-\infty}=r\,\vec{\phi}\cdot\vec{h} =
\frac{2\pi}{e^2}\,\vec{h}\cdot\vec{\phi} \nn\ee
where, in the last equation, we have used the expression for $r$
given in equation \eqn{r}. Comparing with the mass of the monopole
calculated in four dimensions \eqn{mmono}, we find the important
equation, \be M_{\rm kink}=M_{\rm mono} \nn\ee Although we will not
describe the details here, it was shown in \cite{misha,vstring},
following the calculations of \cite{nick}, that this equation
continues to hold when all quantum effects are taken into account:
note that the left-hand-side is computed in a $d=1+1$ dimensional
theory, while the right-hand-side is computed in a $d=3+1$
dimensional theory. The vortex gives a quantitative correspondence
between the theories in different dimensions.

\para
It is simple to see that the kink carries the quantum numbers of the
monopole. It interpolates between two vacua \eqn{vac} on the vortex
worldsheet which, in $d=3+1$ dimensional spacetime, means it joins
two vortex strings with their magnetic field embedded in different
components of the gauge group \eqn{bstar}. The kink can be thought
of as absorbing the magnetic field from the left, and splitting out a
different magnetic field to the right. In other words, it is a
source for the magnetic field with charge $\vec{h}$: it is a
magnetic monopole.

\para
For multiple $k$ strings, the $U(k)$ gauge theory describing 
their dynamics was described in \cite{vib} and 
Section \ref{sc:vortices}%4
. When $k<N_c$, 
there exist vacua of the worldsheet theory in which the adjoint 
chiral multiplet has vanishing vev, and the strings are localised on
top of each other in space. At this point, the kinks on the string
are identical to those discussed in Section \ref{sc:domain-wall} %3 
and their zero modes
are given by the computation on the Appendix B, summarised in
\eqn{final}. As we already noted, the classification and zero mode
structure of domain walls is entirely analogous to that of magnetic
monopoles. Now the physical reason behind this similarity is clear
and is the slogan of this section: the kinks on the vortex string
{\it are} magnetic monopoles.

\para
Finally, one could also ask about the Euclidean vortex on the 
Euclidean vortex string. 
This corresponds to a Yang-Mills instanton trapped inside 
the vortex string \cite{vstring}. 
An explicit realization of this idea was recently obtained 
and the relation between the monopole in the Higgs phase 
and the instanton in the Higgs phase was clarified \cite{EINOS}.

\section{D-Branes}
\label{sc:d-branes}

As we have seen, our theory contains both domain walls and vortex
strings. In fact, it was noticed some years ago that the domain
walls are D-branes for the vortex strings: the vortices may
terminate on the wall where their end is electrically charged under
the $U(1)$ gauge field on the domain wall worldvolume. This was
first shown in \cite{dbrane} for abelian theories in the limit
$e^2\rightarrow\infty$ where explicit solutions were found. The
generalisation to finite gauge coupling was explored in detail by
Shifman and Yung in \cite{shif,shif2} and the solutions were
examined in various limits. Solutions in non-abelian theories in the
$e^2\rightarrow\infty$ limit were discussed in \cite{allquarter}. We
note also that D-branes in the field theoretic context have been
shown to exist in other semi-classical models \cite{mored} and also
arise as a strong coupling phenomenon in which QCD flux-tubes can
end on domain walls \cite{evenmore}.

\para
In this section, we will explore several properties of the 
D-brane system and point out the existence of a new 
contribution to the BPS energy density which was overlooked 
in previous studies: a negative binding energy where the 
string attaches to the domain wall. It is unusual to encounter 
negative contributions to central charges, but we shall argue 
that it is perfectly sensible in this setting. 

\subsubsection*{\it D-branes}

 It turns out that the equations describing a vortex string
ending on a wall are identical to those describing confined
monopoles (\ref{boggy1}) and \eqn{boggy}. 
In fact, these equations were first 
discovered by Shifman and Yung in context of abelian field theoretic
D-branes \cite{shif}. The different solutions follow from imposing
different boundary conditions: for D-branes we require that both
$[{\rm Tr}\phi]^{+\infty}_{-\infty}$ and $\int \Tr B$ are
non-vanishing.

\para
Recall from Section \ref{sc:domain-wall} %3 
that we have $N_f!/N_c!(N_f-N_c)!$ vacua,
providing a large selection of domain walls. Recall also from
Section \ref{sc:vortices} %4 
that in each vacua we have $N_c$ different types of
vortices. Thus the first question we ask is simple\footnote{For the
case of $U(2)$ gauge group with four flavours, this question was
answered in \cite{shif2} in a non-generic case with degenerate
masses. Our answers agree in the relevant limit.}: Which type of
vortex can end on which type of domain wall? Since we have shown
that any system of domain walls can be decomposed into a number of
elementary domain walls labelled by a simple root of
$\Lambda_R(su(N_f))$, let us restrict attention to such an
elementary object, labelled by
%%%%%%%%%%%%%%%%%%%%%%%%%%%%%%%%%%%%
\begin{figure}[htbp]
\begin{center}
\epsfxsize=4.0in\leavevmode\epsfbox{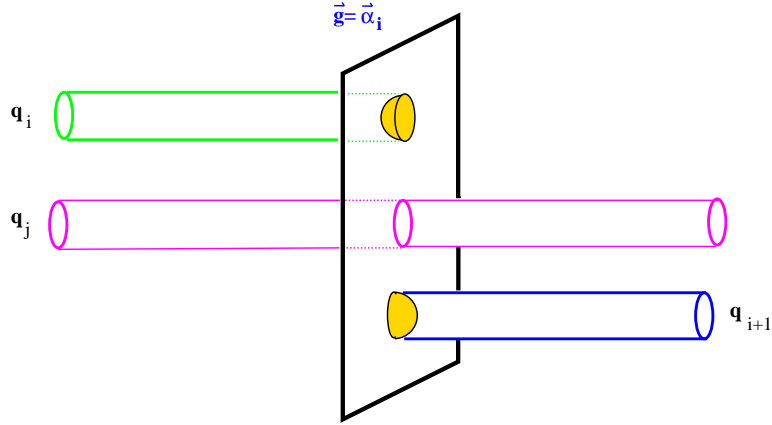}
\end{center}
 \caption{\small %Figure 5: 
For the $\valpha_i$ elementary domain wall, the $q_i$
string may end on the left and the $q_{i+1}$ string may end on the
right. All $q_j$ strings, for $j\neq i,i+1$ exist in both left and
right vacua and pass right through the domain wall. The nodes represent the
finite binding energy.}
\label{fig:vortex-elem-wall}
\end{figure}
%%%%%%%%%%%%%%%%%%%%%%%%%%%%%%%%%%%%%%%%%%%%%%%%%%%%%%%%%%%%%%%%%%%%%%%%%%%%%%%%
%
\be \vec{g}=\valpha_i=(0,\ldots,1,-1,\ldots,0) \nn\ee
where the entry ``$1$'' sits in the $i^{\rm th}$ position. This
means that the vacuum $\tilde{\Sigma}$ to the left of the domain
wall differs from the vacuum $\Sigma$ to the right by a single
element: $i\in\tilde{\Sigma}$ but $i\notin{\Sigma}$, while $(i+1)\in
{\Sigma}$ but $(i+1)\notin\tilde{\Sigma}$. Now we can state our
result: from the vacuum on the left, only the vortex associated to
$q_i$ can end on the domain wall; from the vacuum on the right only
the vortex associated to $q_{i+1}$ is allowed to end. We depict this
in Figure \ref{fig:vortex-elem-wall}. 
In fact, it is simple to see this result and requires
only a quick inspection of the relevant Bogomoln'yi equations 
from (\ref{boggy1}) and 
\eqn{boggy}
\be {\cal D}_3 q_i=(\phi-m_i)q_i \nn\ee
Since we are dealing with an elementary domain wall, only a single
diagonal component of $\phi$ (let's call it $\phi_a$) picks up a
profile such that $\phi_a\rightarrow m_{\tilde{\sigma}(a)} = m_i$ as
$x^3\rightarrow-\infty$, while $\phi_a\rightarrow m_{\sigma(a)} =
m_{i+1}$ as $x^3\rightarrow+\infty$. But, in the vacuum
$\tilde{\Sigma}$, all vortices other than that built around $q_i$
have their winding in the different part of the gauge group and so
it cannot change as we move in the $x^3$ direction and cross the
domain wall.
%%%%%%%%%%%%%%%%%%%%%%%%%%%%%%%%%%%%%%%%%%%%%%%%%%%%%%%%%%%%%%%%%%%%%%%%%
\begin{figure}[htbp]
\begin{center}
\epsfxsize=6.0in\leavevmode\epsfbox{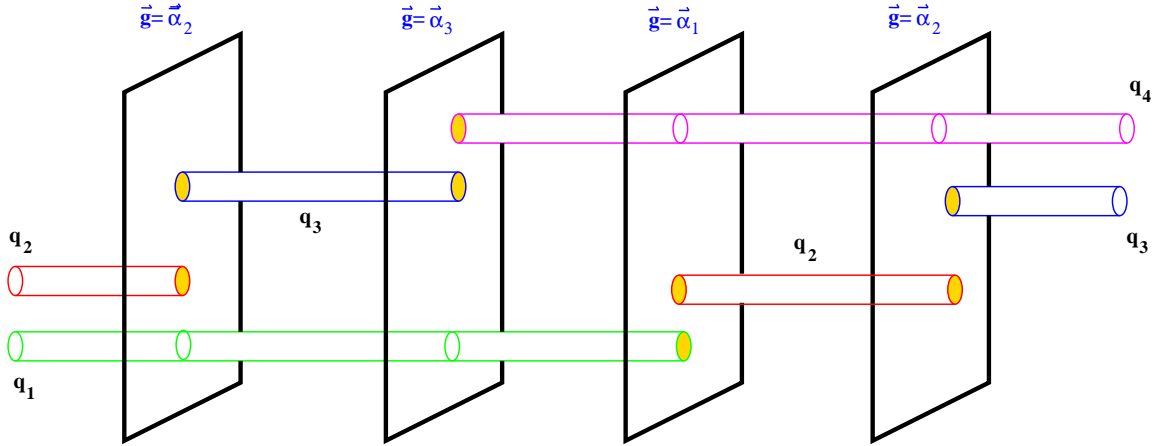}
\end{center}
 \caption{\small %Figure 6: 
Which string on which wall: the allowed
string ends for the $\vec{g}=\valpha_1+2\valpha_2+\valpha_3$ domain
wall system in the $U(2)$ theory with $N_f=4$ flavours.}
\label{fig:allowed-vortices}
\end{figure}
%%%%%%%%%%%%%%%%%%%%%%%%%%%%%%%%%%%%%%%%%%%%%%%%%%%%%%%%%%%%%%%%%%%%%%%%%%%%%%%%

\para
The rule for which strings can end on which domain walls is invariant under permutations
of the domain wall ordering (when allowed). For example, in 
Figure \ref{fig:allowed-vortices} we show
which strings can end on the maximal domain wall system of the
$U(2)$ gauge theory with $N_f=4$. As the inner two domain
walls pass through each other, the strings ending on them are unaffected.

\para
Notice in particular that the rules described above allow a vortex string to be
suspended between adjacent elementary walls {\it only if} the walls cannot pass
each other. For example, in Figure \ref{fig:allowed-vortices} 
we see that it is not possible to suspend a
string between the two inner walls $\valpha_1$ and $\valpha_3$. This means we do not
have to ask the question about what happens to these finite segments of string
as the domain walls change position. However, 
in Section \ref{sc:interaction} %7 
we will discuss a slight
variant of this set-up where this situation does arise.

\para
Finally, let us mention that the simple picture of a rigid string
ending on a rigid domain wall as shown in the figures is not really
accurate. In fact, the string pulls the domain wall and, if the
number of strings ending on the left and right is not equal, the
domain wall bends asymptotically in a logarithmic fashion
\cite{dbrane,shif}. This is familiar behaviour for co-dimension two
objects (for example: the fundamental string ending on a D2-brane,
or the D4-brane ending on an NS5-brane in IIA string theory at
finite coupling). This logarithmic bending will play a crucial role
when we come to study the binding energy.

\subsubsection*{\it Binding Energy}

%%%%%%%%%%%%%%%%%%%%%%%%%%%%%%%%%%%%
\begin{figure}[htbp]
\begin{center}
\epsfxsize=3.5in\leavevmode\epsfbox{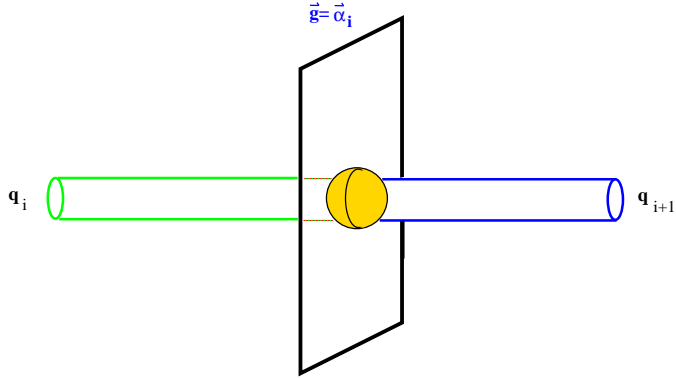}
\end{center}
 \caption{\small %Figure 7: 
When colinear strings end, a binding energy equivalent to a
negative mass monopole appears in the middle.}
\label{fig:binding-energy}
\end{figure}
%%%%%%%%%%%%%%%%%%%%%%%%%%%%%%%%%%%%%%%%%%%%%%%%%%%%%%%%%%%%%%%%%%%%%%%%%%%%%%%%
Now let us turn to the binding energy, a negative contribution to
the total energy of the system. Since this binding energy occurs for
BPS solitons, it can be easily seen by examining the central charge
or, equivalently, the topological charge left after completing the
square in the Hamiltonian. We saw from equation \eqn{finalbog} that
we are left with three such terms: the wall charge, the vortex
charge and the monopole charge. The binding energy is associated
with the latter; more specifically, and rather curiously, it is one
half of a negative mass BPS monopole. The field configuration connecting 
the string to the wall is known in the condensed matter literature 
as a boojum.
%\footnote{This appears to be S-dual to the original proposal
%of \cite{lewis}: ``The squark was a boojum you see''.}

\para
The sign of the binding energy is fixed by the requirement that the tensions of 
the domain walls and vortex strings are positive. From the completion of the 
square in \eqn{finalbog}, we see that this requires us to pick signs for 
$\eta$ and $\epsilon\eta$ respectively; the sign of the monopole central charge 
cannot then be altered and is 
fixed to be $\epsilon$ as in \eqn{mono}. We will now show that the 
monopole contributes a negative energy. We first fix our choice of BPS domain walls 
and BPS vortex strings by choosing $\epsilon=\eta=+1$ in \eqn{finalbog}. 
Start by considering a wall with two vortex strings ending on
it; one from the left and one from the right. 
Let's start with the strings colinear as illustrated 
in Figure \ref{fig:binding-energy}. 
Calculating the contribution to the energy from
the monopole central charge, we have

\be
E_{\rm binding}=- \frac{1}{e^2}\int dx^1dx^2\,\left[\Tr(\phi
B_3)\right]^{+\infty}_{-\infty} =-\frac{2\pi}{e^2}\,(m_i-m_{i+1})<0
\label{twobooj}\ee
%
%%%%%%%%%%%%%%%%%%%%%%%%%%%%%%%%%%%%
\begin{figure}[htbp]
\begin{center}
\epsfxsize=4.0in\leavevmode\epsfbox{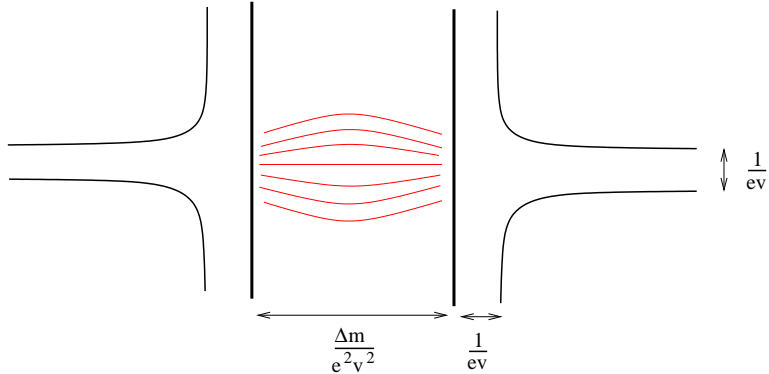}
\end{center}
 \caption{\small %Figure 8: 
Colinear vortex strings ending on the domain wall. In
the centre of the domain wall, the magnetic flux is free to spread
out.}
\label{fig:center-of-wall}
\end{figure}
%%%%%%%%%%%%%%%%%%%%%%%%%%%%%%%%%%%%%%%%%%%%%%%%%%%%%%%%%%%%%%%%%%%%
where the contribution comes from the different values 
adopted by $\phi$ to the left and right of the domain wall,  
and is non-vanishing 
despite the fact that $B_3$ remains constant at left and 
right infinity. Notice that even though this is a finite contribution 
to the energy, we can cleanly separate it from the infinite 
contribution of the vortex strings and the (infinite)${}^2$ 
contribution of the domain wall since it is the only object whose 
energy depends parametrically on the gauge coupling $e^2$.
Usually when we encounter a negative central charge, it means 
we have computed the contribution for an anti-soliton instead of a 
soliton; we should change some minus signs around and
find the correct, positive mass for the BPS state. 
However, this is not the case here.
The minus signs are fixed by requiring that the domain wall 
and vortex contributions are positive: $T_{\rm wall} >0$ 
and $T_{\rm vortex}>0$. 
The negative contribution to the binding energy \eqn{twobooj} 
then follows. The same result can be easily seen to hold for other choices of 
$\epsilon,\eta=\pm 1$.

\para
How should we interpret the binding energy \eqn{twobooj}? We can
find a heuristic explanation by looking at the profile of the domain
wall in the limit $ev \ll \Delta m$ where $E_{\rm binding}$ becomes
large. This profile was drawn in Figure \ref{fig:profile}. 
Now consider what it
looks like with two strings attached. The two outer layers of the
domain wall are the same width as the vortex string and within both
objects one of the Higgs fields $q$ drops exponentially to zero.
This shared property makes it natural for the vortex string to
smoothly merge into the outer layer of the domain wall. But what
occurs in the middle layer? The magnetic flux $\Tr\,\int
dx^1dx^2\,B_3$ must be conserved in the middle of the domain wall.
But here the flux is free to spread out. The BPS soliton must 
therefore look like Figure \ref{fig:center-of-wall}. 
The binding energy can be understood
as the energy difference between the spreading flux in the middle
layer, shown in Figure \ref{fig:center-of-wall}, 
and the energy contained in a collimated
vortex tube of the same length. The latter has already been
accounted for in the topological charge $-v^2\Tr\,\int B_3$, but
this is overcounting; the binding energy remedies this. To see that
the parametric scaling agrees, note that the vortex tube stretched
across the inner layer would have energy
\be
E_{\rm vortex}\sim T_{\rm vortex}\frac{\Delta m}{e^2v^2}\sim \frac{\Delta m}{e^2}
\ee
while the actual energy from the flux can be readily estimated. 
If, for some value of $x^3$, the flux spreads over an area 
$A$ then the magnetic field scales as $B_3\sim 1/A$. 
The effective tension (energy density per unit $x^3$ length) 
of the flux tube in that region is 
\be 
{\cal E}_{\rm flux}\sim \frac{1}{e^2}
\,\Tr\,\int\,dx^1dx^2\, B_3^2\ 
\sim \frac{1}{e^2 A} 
\ee
When the flux is confined in the vortex, it spreads over an 
region $A\sim 1/e^2v^2$, resulting in the vortex tension 
$T_{\rm vortex}\sim v^2$. 
Inside the domain wall, flux is unconfined and 
it spreads over a greater region  $A\sim \alpha/e^2v^2$ 
for some $\alpha > 1$. We therefore have the 
expectation
\be
E_{\rm binding}
=({\cal E}_{\rm flux}-T_{\rm vortex}) A 
%{\Delta m \over e^2 v^2} =E_{\rm flux} - E_{\rm vortex} 
\sim \left(\frac{1}{\alpha}-1\right) \frac{\Delta m}{e^2v^2} 
\sim - \frac{\Delta m}{e^2} < 0
\ee
which is indeed confirmed by the central charge \eqn{twobooj}. 

\para
The phenomenon of a negative BPS binding energy has been encountered 
previously. Wess-Zumino models in $d=3+1$ admit BPS spoke-like solutions 
consisting of multiple domain walls meeting at a junction. The supersymmetry 
algebra was shown to include a central charge for the $d=1+1$ intersection of the 
domain walls \cite{DWJ} following earlier work of \cite{shifchib}. An exact solution 
for the junction of three walls was constructed and the new central charge turned out 
to give a negative contribution to the energy \cite{OINS,tervel}. As with our 
situation, the central charge has the interpretation of 
a BPS binding energy.

%\para
%When the strings are colinear, it is clear from the discussion of
%the last section that we can interpret the object \eqn{twobooj} as a
%confined monopole. The only novel feature is that it is restricted
%to lie at the intersection of the string and domain wall. This must
%be the case because the $q_{i+1}$ string does not exist to the left
%of the domain wall, while the $q_i$ string does not exist to the
%right. So the monopole cannot slide along the vortex string.

\para
So much for colinear vortices. What is the binding energy for a single
string ending on a domain wall? The answer, as we shall now argue, is
that a single string has binding energy
\be E_{\rm binding} = - \frac{1}{2}\, \frac{2\pi}{e^2}\,(m_i-m_{i+1}) =
-\frac{1}{2}M_{\rm mono} \label{booj}\ee
At first glance, this seems rather unlikely! How can this fractional (negative)
mass arise from the topological charge given the well-known
quantisation conditions? To see how %it 
this feat is achieved,
consider a situation where %there 
the vortex string
ends from the left. When evaluated at $x^3\rightarrow -\infty$,
the topological charge gives
\be \frac{1}{e^2}\left.\int\,dx^1dx^2\,\Tr(\phi
B_3)\right|_{-\infty} = -\frac{2\pi m_i}{e^2} \label{minus}\ee
At right infinity it naively appears that there is no contribution
since there are no vortex strings there. However, this is not
correct because, as we mentioned previously, the domain wall bends
logarithmically away from the terminating string. Denoting the
center of mass of the domain wall as $x^3_{(0)}$, the asymptotic
profile of the D-brane is $x^3_{(0)}\sim \log |z|$
\cite{dbrane,shif}. Crucially for us, this means that the magnetic
flux is deposited on the domain wall by the vortex string, and then
carried to $x^3\rightarrow +\infty$ by this logarithmic bending,
resulting in a contribution to the topological charge. Since the
domain wall interpolates from $\phi_a = m_i$ at $x^3=-\infty$ to
$\phi_a=m_{i+1}$ at $x^3=+\infty$, in the center of the domain wall
the scalar field takes the average value
$\phi_a=\ft12(m_i+m_{i+1})$. Thus the contribution to the charge at
right infinity is
\be -\frac{1}{e^2}\left.\int dx^1dx^2\,\Tr(\phi
B_3)\right|_{+\infty} = +\frac{2\pi}{e^2}\,
\frac{1}{2}\,(m_i+m_{i+1}) \label{plus}\ee
Adding together the two contributions \eqn{minus} and \eqn{plus}, we
find that the biding energy of the string ending on the wall is
indeed given by half the mass of the monopole.
%%%%%%%%%%%%%%%%%%%%%%%%%%%%%%%%%%%%
%\begin{figure}[htbp]
%\begin{center}
%\epsfxsize=3.5in\leavevmode\epsfbox{splitting.eps}
%\end{center}
% {\small Figure 8: If the string ends are not lined up on the wall,
%the monopole cleaves in two. Each half is the boojum.}
%%\label{profile}
%\end{figure}
%%%%%%%%%%%%%%%%%%%%%%%%%%%%%%%%%%%%%%%%%%%%%%%%%%%%%%%%%%%%%%%%%%%%%%%%%%%%%%%%
%\para
%The arguments of this section apply equally well to abelian $U(1)$
%gauge theories. Indeed, in restricting to an elementary domain wall
%we are essentially focussing attention on a single $U(1)\subset
%U(N_c)$ subgroup of the non-abelian gauge group. Thus the boojums --
%which we have just seen are half-monopoles -- exist even in abelian
%theories, even though these do not have the topology to support
%magnetic monopoles in the Coulomb phase.

\para
Let us comment that the presence of the binding energy
differentiates these field theoretic D-branes from those of type II
string theory. Naively it appears that stretched strings become
massless, and then tachyonic, as domain walls approach and overlap.
Were they to truly become unstable, the result would be to Higgs
the axial $U(1)$ on the worldvolume under which they are charged,
thereby reducing the number of collective coordinates
\cite{schmaltz}. Since the index theorem of Appendix B shows no sign
of this behaviour, it seems that the domain walls cannot approach
close enough to allow the string to become tachyonic. (Recall that
penetrable domain walls, which can pass through each other, cannot
have strings stretched between them).

\para
More interesting is the possibility that the stretched strings become
exactly massless, resulting in non-abelian symmetry enhancement.
This was conjectured to occur in \cite{shif2} in the non-generic
case of degenerate masses, although the non-abelian nature was
exhibited only at the linearised level. 
%Note that in this limit the binding energy vanishes since 
%$m_i=m_{i+1}$. 
%If this is true, it should show up as a 
%singularity in the corresponding domain wall moduli space. 
%To our knowledge, this metric has been computed only in the 
%abelian case in the limit $e^2\rightarrow \infty$ (in which 
%the binding energy also vanishes) where it was shown to be 
%smooth\cite{tong}. 
It would be interesting to examine the domain wall
moduli space in the non-abelian theory to find evidence for
non-abelian symmetry enhancement.

\subsubsection*{\it An Aside: Boojums in Superfluid ${}^3$He}

The term boojum was introduced to physics by Mermin in the context
of superfluid ${}^3$He \cite{mermin}. If ${}^3$He-A is placed in a
closed, spherical container, the boundary conditions imposed at the
surface require that the order parameter becomes singular at a
point. This point -- the boojum -- emits a vortex texture into the
bulk of the fluid. More pertinent to the present discussion, 
boojums also occur on the AB-interface \cite{ab}. This interface is 
a domain wall separating the two phases of superfluid ${}^3$He. A
global vortex in the ${}^3$He-B phase may end on the domain wall,
where it terminates on a boojum. The binding energy of our system 
can be thought of as a negative mass boojum. In ${}^3$He, as in our case, 
the boojum is a close relative of the hedgehog (or monopole). For a
nice introduction to this subject, see the entertaining book
\cite{volovik}. Boojums are also observed in other condensed matter
systems, including cholesteric and smectic liquid crystals
\cite{smeg}.

\subsubsection*{\it The View from the Domain Wall}

In Section \ref{sc:domain-wall}%3
, we reviewed the low-energy dynamics on an 
elementary domain wall. 
From this perspective, the vortex strings are rather 
simple to see: they correspond to spike-like solutions to the 
low-energy effective action \cite{dbrane,shif}, analogous 
to the ``BIon'' solutions of the Born-Infeld action for 
D-branes in string theory \cite{bion}. 
For $k_R$ strings ending from the right at position 
$z^{R}_p$ (recall $z=x^1+ix^2$) and $k_L$ strings ending 
from the left at positions $z^{L}_q$, the solution on the 
domain wall worldvolume is 
\be \frac{2T_{\rm wall}}{v^2}\,X(z) =
\log\left(\frac{\prod_{q=1}^{k_L}(z-z^{L}_q)}
{\prod_{p=1}^{k_R}(z-z^{R}_p)}\right) \label{xsol}\ee
where $X=x^3_{(0)}+i\theta$ was defined in equation \eqn{dothedual}.
Note that the domain wall bends logarithmically as $|z|\rightarrow
\infty$ unless $k_L=k_R$. The holomorphic property of this solution
is sufficient to ensure that it preserves $1/2$ of the four
supercharges on the domain wall worldvolume, corresponding to a
$1/4$-BPS state from the $d=3+1$ dimensional perspective.

\para
Consider a single vortex string ending from the left. As described
in detail in \cite{dbrane} and \cite{shif}, as we travel once around
infinity $z\rightarrow e^{2\pi i}z$, the phase angle on the domain
wall worldvolume shifts: $\sigma\rightarrow \sigma +2\pi$. In other
words, the vortex string is a global vortex for $\sigma$. Switching
to the dual field strength $F$ on the worldvolume of the domain wall
\eqn{donthedual}, we see that the vortex string is electrically
charged.

\para
The low-energy action on the domain wall given in \eqn{dual} is
valid only at suitably large length scales. For example, it
certainly does not hold at length scales comparable to the width of
the domain wall which, as we reviewed in 
Section \ref{sc:introduction}%1
, varies from $1/\Delta m$ to $\Delta m/e^2v^2\gg 1/\Delta m$ as we dial the
dimensionless parameter $ev/\Delta m$ from large to small. This
limitation of the domain wall dynamics can be seen in the solution
\eqn{xsol} which describes an increasingly narrow spike as
$z\rightarrow z_q^L$ or $z\rightarrow z_p^R$ whereas, in reality,
the vortex string stabilises at a width $1/ev$. This breakdown of
the domain wall dynamics at short distances also manifests itself in
another way. It is a simple matter to evaluate the energy of the
solution \eqn{xsol} using the domain wall theory \eqn{firstx}. One
finds $k_L+k_R$ ultraviolet divergences which can be interpreted as
the infinite vortex strings. For $k_L\neq k_R$, there is also an
infra-red divergence arising from the asymptotic bending of the
domain wall. However, there is no hint of the binding energy.
Indeed, we have seen that this binding energy occurs over a distance
comparable to width of the domain wall and so might suspect that the
worldvolume theory is too coarse to capture it. However, given the
BPS nature of the binding energy there may exist a formulation in
which it is seen by the domain wall dynamics. We leave this as an
open question.

\section{Rules of Interaction}
\label{sc:interaction}

In this final section we would like to discuss further rules of
interaction between the various solitons described above. We will
now consider all possible combinations of objects -- walls, strings,
and monopoles -- and ask how they join together and under
what circumstances different objects may pass through each other.
Our analysis is based purely on the energetics of the BPS solitons
and we find a unique set of rules {\it under the assumption} that
there exist solutions to the Bogomoln'yi equations 
(\ref{boggy1}) and \eqn{boggy} 
corresponding to far-separated solitons of various types. While such
a ``no-force'' condition is a familiar consequence of the BPS
property of solitons, counterexamples are known (see for example
\cite{multi}) and it is important to stress that while our crude
analysis results in a consistent and unique picture of the
interactions, it  falls short of proving that the rules we describe
occur dynamically. Further work attempting to demonstrate the
existence of the solutions is in progress.

\para
Firstly consider solitonic configurations which involve both 
domain walls and confined monopoles. 
The domain walls are $1/2$-BPS; the confined monopoles are 
$1/4$-BPS. When combined, it turns out that 
the configuration still preserves $1/4$ of the 
supersymmetry \cite{allquarter}, 
rather than $1/8$ as one might expect. 
This can be seen when completing the square in \eqn{finalbog} 
by noting that one has only two choices of $\pm$ signs, $\epsilon$ 
and $\eta$ rather than three. Fixing $\eta$ and $\epsilon\eta$ determine 
the half of the supercharges preserved by the domain wall and vortex respectively. 
The two may coexist peacefully, preserving $1/4$ of the supersymmetry. We have no 
further $\pm$ signs to play with and only monopole for which $M_{\rm mono}>0$ 
preserve this remaining supersymmetry.  
%%%%%%%%%%%%%%%%%%%%%%%%%%%%%%%%%%%%
\begin{figure}[htbp]
\begin{center}
\epsfxsize=2.5in\leavevmode\epsfbox{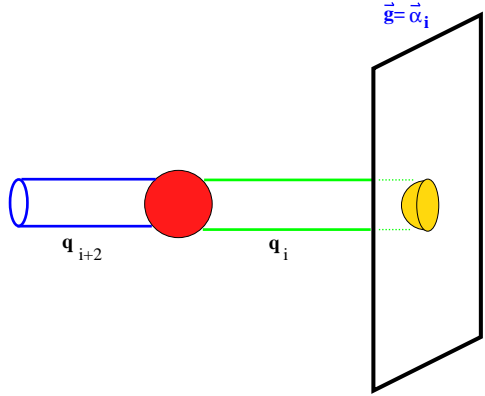}
\end{center}
 \caption{\small %Figure 9: 
A configuration allowed by supersymmetry. 
Is the position
of the monopole a modulus, or is it attracted or repelled 
by the wall?}
\label{fig:SUSY-config}
\end{figure}
%%%%%%%%%%%%%%%%%%%%%%%%%%%%%%%%%%%%%%%%%%%%%%%%%%%%%%%%%%%%%%%%%%%%%%%%%%%%%%%%

\para
In terms of gamma matrices, this discussion is somewhat 
simpler~\cite{allquarter2}. 
The domain wall and vortex preserve $1/2$ of the 
supercharges, determined by the positive eigenspace of a 
suitably defined gamma matrix: $\xi=\gamma_{\rm wall}\xi$ and $\xi=\gamma_{\rm vortex}\xi$. 
The fact that these solitons can be mutually $1/4$ BPS is the statement that 
$[\gamma_{\rm wall},\gamma_{\rm vortex}]=0$. Finally, the monopole breaks no 
further supersymmetries because its gamma matrix is related to the other two: 
$\gamma_{\rm mono}=\gamma_{\rm wall}\gamma_{\rm vortex}$.

\para
%Let us stress that both signs of vortex magnetic flux are 
%allowed, but they preserve different $1/4$ of supercharges, 
%which we shall refer as BPS for $\int\,\Tr\,B_3<0$ 
%and anti-BPS for $\int\,\Tr\,B_3>0$. 
%Therefore monopole charge should also be positive for 
%the $1/4$ BPS states and negative for the $1/4$ anti-BPS 
%states. 
%In order to make arguments simpler, we will consider mostly 
%the case of $1/4$ BPS states with $\int\,\Tr\,B_3<0$ 
%and $\int\,\partial_\alpha(\phi B_\alpha)<0$ rather than 
%anit-BPS states, unless stated otherwise. 

The fact that $1/4$ of the supersymmetry is preserved has an 
important consequence: in the presence of a domain wall, a 
confined monopole may only be oriented one way in the $x^3$ 
direction and still preserves supersymmetry. This follows immediately 
from the $\eta$ and $\epsilon$ assignments in \eqn{finalbog}. If we 
choose BPS domain walls with $\eta=+1$ and BPS vortex strings 
with $\epsilon=+1$ (corresponding to ${\rm Tr}\int B_3 <0$) 
then we already saw in Section 5 that the Bogomoln'yi equation 
\eqn{boggy1} and \eqn{boggy} require that a $q_k$ string should be 
emitted to the left and the $q_j$ string to the right of the wall, 
if $k>j$. 
%Domain walls separate two domains of different vacua. 
%Let us recall the ordering of vacua in our choice of 
%$1/2$ BPS projection for walls. 
%For each color component, the flavor of the vacuum should 
%increase as $x^3$ increases. 
%If flavors $k$ and $j$ are involved in forming a wall, 
%it can emit both a $q_k$ string and a $q_j$ string. 
%Consequently the 
To see that this is the orientation compatible with domain 
walls, simply note that the Bogomoln'yi equations for the 
confined monopole (\ref{boggy1}) and 
\eqn{boggy} contain within them the BPS 
domain wall equations \eqn{bogwallone} and \eqn{bogwalltwo}. 
While the condition above was derived for BPS vortices defined 
as $\int\,\Tr\,B_3<0$, it can be easily checked that the same 
holds for BPS anti-vortices with $\int\,\Tr\,B_3>0$. 

%%%%%%%%%%%%%%%%%%%%%%%%%%%%%%%%%%%%
\begin{figure}[htbp]
\begin{center}
\epsfxsize=5.8in\leavevmode\epsfbox{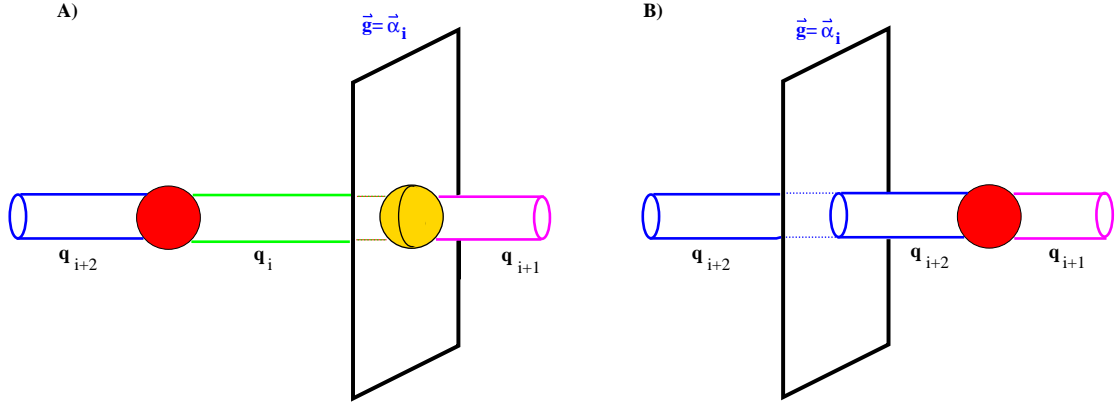}
\end{center}
 \caption{\small %Figure 10: 
A monopole passes through the domain wall, soaking up the
negative binding energy in the process.}
\label{fig:passing-monopole}
\end{figure}
%%%%%%%%%%%%%%%%%%%%%%%%%%%%%%%%%%%%%%%%%%%%%%%%%%%%%%%%%%%%%%%%%%%%%%%%%%%%%%%%
\para
Now we understand which orientation of confined monopoles is 
compatible with the orientation of vortices, %domain walls, 
we can try to build a soliton configuration consisting of 
walls, vortices and monopoles. 
These putative BPS configurations 
should %would once more 
satisfy the Bogomoln'yi equations \eqn{boggy} with 
the appropriate topological charges arising from the boundary 
conditions. 
For example, if the left vacuum $\tilde{\Sigma}$ 
supports vortex strings of the $q_{i+2}$ and $q_i$ type, 
a possible BPS configuration would be a $q_{i+2}$ string 
entering from the left, transforming into a $q_i$ string 
by the presence of a monopole, and subsequently ending on 
a $\vec{g}=\valpha_i$ domain wall. 
If this monopole were in the Coulomb phase it would be able 
to decompose into two fundamental monopoles but, in the 
vacuum $\tilde{\Sigma}$, this is not possible since there 
is no $q_{i+1}$ vortex string to take away the intermediate 
component magnetic flux. 
(By definition of the $\vec{g}=\valpha_i$ domain wall, only 
the right vacuum supports such a string). 
The configuration is shown in Figure \ref{fig:SUSY-config}. 
We stress that we do not know whether the distance between 
the monopole and domain wall is a modulus of such a solution 
as suggested by the BPS bound and a naive application of the 
``no-force'' folk theorem. We would be interested in knowing 
if this is indeed the case.

\para
Let us now ask whether monopoles are permitted to pass through
domain walls. Obviously they can only pass through if there is a
colinear string waiting on the other side onto which they can slide.
Suppose that we have a $\vec{g}=\valpha_i$ domain wall in which both
left and right vacua support $q_{i+2}$ vortex strings. We can
reconsider the situation of Figure \ref{fig:SUSY-config}, 
but now where there is a
$q_{i+1}$ string attached to the right as shown in 
Figure \ref{fig:passing-monopole}A.
Energetically, it is now possible for the monopole to pass through
the domain wall. As it does so, it absorbs the negative binding
energy, leaving behind a $q_{i+2}$ string which penetrates the
domain wall unopposed. In the right vacuum $\Sigma$, sits a monopole
of lower charge, connecting the $q_{i+2}$ string to the $q_{i+1}$
string. It's mass is lower than the original monopole by an amount
equal to the binding energy. This final configuration is shown in
Figure \ref{fig:passing-monopole}B.

%%%%%%%%%%%%%%%%%%%%%%%%%%%%%%%%%%%%
\begin{figure}[htbp]
\begin{center}
\epsfxsize=4.8in\leavevmode\epsfbox{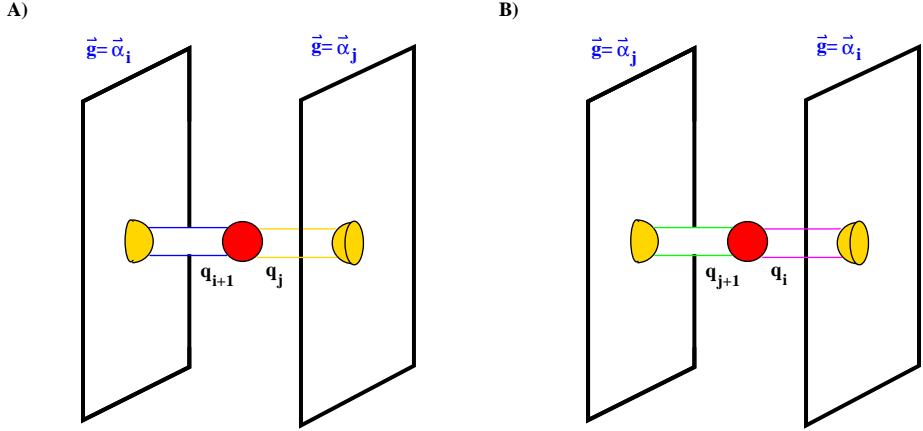}
\end{center}
 \caption{\small %Figure 11: 
Crushing the monopole. The two configurations with the 
domain walls interchanged carry the same asymptotic charge, 
but only the configuration on the left preserves supersymmetry.}
\label{fig:crush-monopole}
\end{figure}
%%%%%%%%%%%%%%%%%%%%%%%%%%%%%%%%%%%%%%%%%%%%%%%%%%%%%%%%%%%%%%%
\para
Another question: what becomes of monopoles that are trapped between
domain walls as the walls approach and pass through each other? The
answer  depends on whether there are further strings extending to
the left and right providing an escape route for the monopole. Let's
first examine the case where there can be no escape. Consider two
penetrable elementary domain walls $\valpha_i$ and $\valpha_j$ with
$\valpha_i\cdot\valpha_j=0$. Assume that $i>j$. (In fact $i>j+1$ since
they are penetrable). We saw in
Section \ref{sc:d-branes} %6 
that a simple vortex
string cannot be suspended between adjacent, penetrable walls. But the
BPS conditions allow us to stretch a string between the walls with a monopole
suspended in the middle. This is shown in 
Figure \ref{fig:crush-monopole}. When the
$\valpha_i$ domain wall is to the left of the $\valpha_j$ domain
wall, as in Figure \ref{fig:crush-monopole}A, 
we see that the $q_{i+1}$ string is to the
left of the $q_j$ string. By our criterion above, this 
ensures that the objects in the configuration are mutually 
BPS for vortex strings satisfying $\int \Tr B_3 < 0$. 
The monopole charge for this configuration has three 
contributions: from the left and right binding 
energies, and from the central monopole.
\be 
\frac{e^2}{2\pi}M_{\rm mono} &=& -\frac{1}{2\pi}
\left.\int\,dx^1dx^2\,\Tr(\phi B_3)\right|^{+\infty}_{-\infty} 
\nn\\ &=& -\ft12 (m_{i}-m_{i+1}) +(m_j-m_{i+1})-\ft12(m_j-m_{j+1})>0  
\label{chargeit}
\ee
Now consider attempting to exchange the position of the 
domain walls. 
What becomes of the monopole and string? 
%Without the monopole and vortex, the two walls are penetrable. 
%Therefore their positions can be exchanged by gradually 
%changing the relative distance (one of the moduli parameters). 
%During the process, the boundary conditions as well as 
%the preserved supercharges are unchanged. 
Since the charge \eqn{chargeit} is measured at infinity, it 
cannot simply disappear. Indeed, there is a configuration carrying 
the correct charge, drawn in Figure 11B. However, to get the 
minus signs right, we require that the strings in this configuration 
are anti-vortices, with ${\rm Tr}\int B_3 > 0$. The BPS monopole 
charge for Figure 11B is given by,
\be 
\frac{e^2}{2\pi}M_{\rm mono} 
= \ft12 (m_{j}-m_{j+1}) +(m_{j+1}-m_{i})+\ft12(m_i-m_{i+1})  
\label{chargeit2}
\ee
which agrees with \eqn{chargeit}. However, from the discussion above 
we see that this configuration breaks all supersymmetry. The orientation 
of the confined monopole is not compatible with the presence of 
domain walls. We therefore conclude that the transition from Figure 11A 
to Figure 11B is not smooth. It seems likely that no solution of the 
form 11B exists. 

\para
It is an interesting phenomenon that a pair of penetrable walls 
can no longer exchange positions once a vortex with a monopole 
stretches between them. This is made more plausible when we 
recall that the vortex string causes the two domain walls 
to bend logarithmically to infinity. In Figure 11A, the 
$\valpha_i$ domain wall bends to left infinity, while in 
Figure 11B it would bend to right infinity. Thus, despite 
appearances, the two configurations are actually separated by 
an infinite distance in field space. Both this, and the supersymmetry 
argument above, point to the existence of a minimal separation between the 
two domain walls in Figure 11A when a confined monopole is hung between them.

%If we accept that the total energy depends on the relative 
%distance between walls, we may question the concept of 
%the relative distance between walls being a modulus in this 
%case, since the energy of a BPS configuration should be 
%determined by topological charges which are fixed once 
%boundary conditions are fixed. 

%Since the charge \eqn{chargeit} is measured at infinity, it 
%cannot simply dissapear. Indeed, there is a configuration carrying 
%the correct charge, drawn in Figure \ref{fig:crush-monopole}B. 
%However, to get the minus signs right, we require that the strings 
%in this configuration are anti-vortices, with $\int\,\Tr\,B_3>0$. 
%The {\it BPS} monopole charge for Figure \ref{fig:crush-monopole}B 
%is given by,
%
%\be \frac{e^2}{2\pi}M_{\rm mono}=\ft12
%(m_j-m_{j+1})+(m_{j+1}-m_i)+\ft12(m_i-m_{i+1}) \nn\ee
%which agrees with \eqn{chargeit}. From the discussion above, 
%we see that this second configuration in Figure \ref{fig:crush-monopole}B 
%breaks supersymmetry: the orientation of the strings emitted by the 
%confined monopole is not compatible with the presence of the 
%domain walls. 
%We therefore conclude that the transition from 
%Figure \ref{fig:crush-monopole}A to 
%Figure \ref{fig:crush-monopole}B cannot be a smooth process. 
%It seems plausible that no BPS solution of the form 11b 
%exists. 

\begin{figure}[htbp]
\begin{center}
\epsfxsize=5.5in\leavevmode\epsfbox{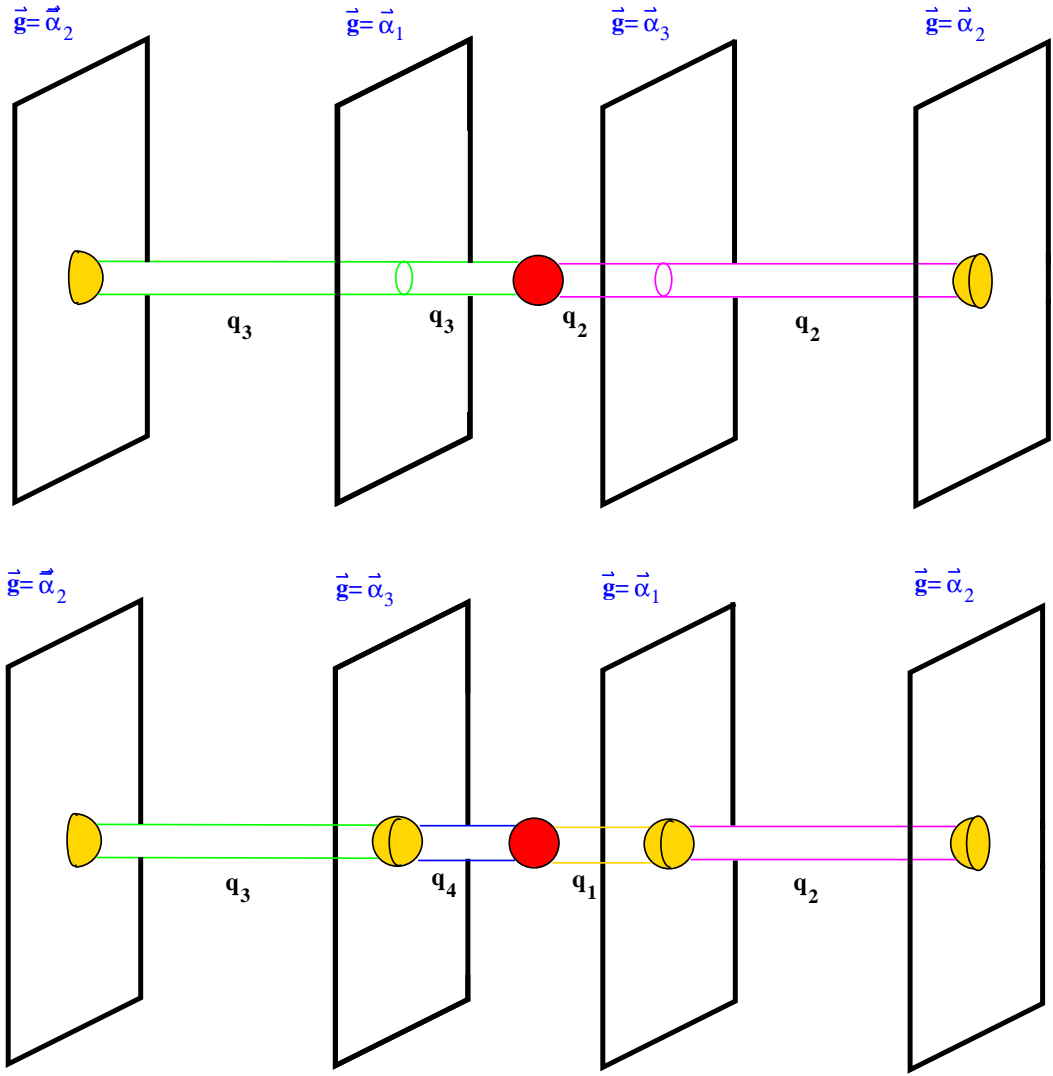}
\end{center}
 \caption{\small %Figure 12: 
A monopole trapped between two domain walls as they pass. In
this case the monopole can escape. The final, supersymmetric,
configuration is depicted in the lower figure after the inner two
domain walls have exchanged positions.}
\label{fig:trapped-monopole}
\end{figure}

\para
Finally, let's turn to an example where the monopole can escape the
crush by returning to the case of four walls in the $U(2)$ gauge
theory with $N_f=4$ flavours. An interesting monopole configuration
is drawn in Figure \ref{fig:trapped-monopole}A. 
We want to know what happens as the $\valpha_1$
and $\valpha_3$ domain walls change position. A configuration with
the same monopole charge and the walls interchanged is drawn in
Figure \ref{fig:trapped-monopole}B. 
This time, and in contrast to the situation in 
Figure \ref{fig:crush-monopole}, 
both configurations preserve the same set of supercharges,
suggesting that the transition is smooth. Indeed, it is also
possible to get from Figure \ref{fig:trapped-monopole}A to 
\ref{fig:trapped-monopole}B by first moving the monopole
out of harms way; the monopole may slide either left or right, past
the $\valpha_1$ or the $\valpha_3$ wall, promoting itself to a
higher charge monopole while leaving behind a negative binding
energy. We can now allow the $\valpha_1$ and $\valpha_3$ domain
walls to pass through each other, and finally slide the monopole
back into the middle vacuum. The end result is that the monopole has
grown to a higher charge object, with its excess energy accounted
for in two binding energies on the $\valpha_1$ and $\valpha_3$
domain walls.
%%%%%%%%%%%%%%%%%%%%%%%%%%%%%%%%%%%%
%%%%%%%%%%%%%%%%%%%%%%%%%%%%%%%%%%%%%%%%%%%%%%%%%%%%%%%%%%%%%%%%%%%%%%%%%%%%%%%%

\newpage
\section*{Appendix A: Gauge Theories and Massive Sigma Models}
\label{sc:NLSM}

As described in \cite{phases}, the long-wavelength limit of the gauge theory
\eqn{ham} is a massive sigma-model with target space given by the Higgs branch
of the gauge theory. Many of the solitons described below are somewhat simpler
in the sigma-model and, indeed, were first discovered in that context.
Here we briefly recall how the relationship to the sigma model works.

\para
The relevant limit is at strong coupling $e^2\rightarrow\infty$. Here the
D-terms written in the second line of \eqn{lag} are imposed as constraints.
Ignoring the mass terms
for now, the moduli space of solutions to this equations is given by the
possible choices of $N_c$ orthonormal vectors in ${\bf C}^{N_f}$. The
$U(N_c)$ gauge transformations act by rotating this basis of vectors
meaning that the space of gauge invariant solutions is given by the
set of $N_c$-dimensional planes in ${\bf C}^{N_f}$. This is simply the
symplectic quotient construction of the Grassmannian $G(N_c,N_f)$. Thus the
Higgs branch is
\be
{\cal M}_{\bf Higgs} \cong G(N_c,N_f)
\label{mhiggs}\ee
(Including the $\tilde{q}$ fields in the analysis, and working with the D-term
of \eqn{lag} results in the cotangent bundle ${\cal M}_{\rm Higgs}=T^\star G(N_c,N_f)$.
However, all soliton solutions live on the zero-section $G(N_c,N_f)$ which is the
sigma-model translation of the fact that we  can ignore the $\tilde{q}$ fields. In a
slight abuse of notation, we will therefore continue to refer to the Grassmannian as
the Higgs branch).

\para
In the absence of the masses, we would simply be left with a sigma model with
target space ${\cal M}_{\rm Higgs}$. The metric on this space descends from
the kinetic terms of the gauge theory via the K\"ahler quotient construction.
We denote this metric as $g_{pq}$ where  $p,q=1,\ldots 2N_c(N_f-N_c)$.
In the presence of the masses, the sigma-model kinetic term is augmented with a
potential given by
\be
V=\sum_{i=1}^{N_f}q_i^\dagger(\phi-m_i)^2q_i
\nn\ee
where $\phi$ is to be taken to solve its equation of motion which, in the strong
coupling limit, is algebraic $\sum_i(\phi-m_i)q_iq_i^\dagger=0$. In fact,
there is a very nice geometrical interpretation for the potential $V$ when
written in terms of the sigma-model. To see this, it is necessary to discuss
the symmetries of the Higgs branch. When the masses are set to zero, the $SU(N_f)$
flavour symmetry has a natural action on ${\cal M}_{\bf Higgs}$ resulting in an
$SU(N_f)$ isometry of the target space. The masses -- and hence the potential $V$
-- break this to $U(1)^{N_f-1}$. Consider the associated $N_f-1$ mutually commuting
holomorphic Killing vectors on ${\cal M}_{\rm Higgs}$. In fact, it will prove useful
to overcount by embedding $SU(N_f)\subset U(N_f)$ in the natural fashion,
resulting in $N_f$ Killing vectors $K_i$ on ${\cal M}_{\rm Higgs}$ satisfying
the constraint $\sum_iK_i=0$. Then it is not hard to show that the potential $V$
can be rewritten as
\be
V=\sum_{i,j=1}^{N_f}\,(m_iK_i^p)\,(m_jK_j^q)\,g_{pq}
\nn\ee
Potentials of this form were first described by Alvarez-Gaum\'e and Freedman
who showed that this is the unique potential allowed by sigma-models with
8 supercharges \cite{agf} (see also \cite{intersect} for a very slight
generalisation).

\para
We see that in the language of the sigma-model, the vacua of the theory
$V=0$ are the fixed points of the $U(1)^{N_f-1}$ action on $G(N_c,N_f)$. On
general grounds (for example, using Morse theory) it is known that there are
$N_{\rm vac}=\chi(G(N_c,N_f))=N_f!/N_c!(N_f-N_c)!$ such fixed points, in agreement with
the gauge theory analysis above.

\section*{Appendix B: The Index Theorem for Domain Walls}
\label{sc:indexth}

In this appendix we perform the calculation of the Callias index theorem in order
to count the zero modes of a domain wall solution interpolating between different vacua.
For domain walls in abelian gauge theories, the index theorem was computed by
K. Lee in \cite{keith} while for domain walls in the sigma-model limit a Morse theory
argument was given in \cite{multi}. Here we give the calculation for the non-abelian
gauge theory and derive the result \eqn{final}.

\para
We let $q$ denote a $N_c\times N_f$ matrix. Similarly, let $m$ denote the $N_f\times N_f$
matrix which is simply $m={\rm diag}(m_1,\ldots,m_{N_f})$. Then the domain wall
Bogomoln'yi equations
can
be written as
\be
{\cal D}_3\phi&=&e^2(qq^\dagger-v^2)  \label{vort1}\\
{\cal D}_3q&=&\phi q-qm
\label{vort2}\ee
We linearise around a background solution and write $\delta\phi=\hat{\phi}$ with
similar notation for other fields. We have
\be
{\cal D}_3\hat{\phi}-i[\hat{A}_3,\phi]&=&e^2(\hat{q}q^\dagger+q\hat{q}^\dagger)\label{bog1}\\
{\cal D}_3\hat{q}-i\hat{A}_3q&=&\phi\hat{q}+\hat{\phi}q-\hat{q}m
\label{bog2}\ee
which are to be supplemented by a gauge fixing condition which naturally comes from
Gauss' law and is taken to be
\be
{\cal D}_3\hat{A}_3-i[\phi,\hat{\phi}]=ie^2(q\hat{q}^\dagger-\hat{q}q^\dagger)
\label{gauss}\ee
We can combine \eqn{bog1} and \eqn{gauss} by introducing the complex adjoint field
$\xi = \phi + iA_3$ in terms of which the equations become,
\be
\partial_3 \hat{\xi}+[\xi^\dagger,\hat{\xi}]-2e^2\hat{q}q^\dagger&=&0
\label{bog3}\\
\partial_3\hat{q}-\xi\hat{q}+\hat{q}m-\hat{\xi}q&=&0
\label{bog4}\ee
We wish to determine the number of solutions to the two 
complex linearised Bogomoln'yi equations \eqn{bog3} and 
\eqn{bog4}. 
We can rescale the fields to absorb $2e^2$ in Eq.~\eqn{bog3} 
without affecting the number of solutions. 
In the following we take $2e^2\rightarrow 1$. 
To make the notation of
matrix multiplication somewhat simpler, it is useful to take the transpose
of these equations. We define $h=q^T$, an $N_f\times N_c$ matrix, and
$\zeta=\xi^T$, an $N_c\times N_c$ matrix. The advantage of working with the
transpose matrices is that our two equations
may now be combined into a single matrix equation,
\be
\Delta \left(\begin{array}{c}\hat{\zeta} \\ \hat{h} \end{array}\right)=0
\ee
where $\Delta$ is an $(N_f+N_c)\times (N_f+N_c)$ square 
matrix operator given by 
\be
\Delta=\left(\begin{array}{cc}\partial_3-\zeta_a^\dagger & -h^\dagger \\ -h &
\partial_3-\zeta_R+m\end{array}\right). 
\ee
The operator $\zeta_a^\dagger$ acts by adjoint action, while the operator
$\zeta_R$ acts by right multiplication (as indeed it must by simply examining
the indices).
The operator $\Delta$ acts on a vector space of 
 $(N_c+N_f)\times N_c$ matrices 
$\psi=(\alpha,\beta)^T$ where the inner product between 
$\psi=(\alpha,\beta)^T$ and $\phi=(\gamma,\delta)^T$  
is defined by 
\be
\left(\phi, \psi\right)
&\equiv &
{\rm Tr}\left[\left( \gamma, \delta \right)
\left(\begin{array}{c}\alpha \\ \beta\end{array}\right)\right] 
\equiv 
\left(\gamma, \alpha\right)+\left(\delta, \beta\right)
\nonumber \\
&\equiv &
\int dx^3 \left(
{\rm Tr}\left[\gamma^\dagger(x^3) \alpha(x^3)\right] 
+{\rm Tr}\left[\delta^\dagger(x^3) \beta(x^3)\right]
\right) .
\label{eq:innerprod}
\ee
Therefore the adjoint operator $\Delta^\dagger$ is given by 
\be
\Delta^\dagger=\left(\begin{array}{cc}-\partial_3-\zeta_a & -h^\dagger \\
-h & -\partial_3 - \zeta_R^\dagger +m\end{array}\right) .
\ee

To proceed, we firstly show that the adjoint operator $\Delta^\dagger$ has no zero modes.
Then the norm of $\Delta^\dagger$ acting on an arbitrary 
$(N_c+N_f)\times N_c$ matrix 
$(\alpha,\beta)^T$ is given by 
\be
|\Delta^\dagger\left(\begin{array}{c}\alpha \\ 
\beta\end{array}\right)|^2 &=&
|\partial_3\alpha+[\zeta,\alpha]+h^\dagger\beta|^2
+|\partial_3\beta+\beta\zeta^\dagger-m\beta+h\alpha|^2
\label{nozero}\\
&=& |\partial_3\alpha+[\zeta,\alpha]|^2 
+ |h^\dagger\beta|^2
+|\partial_3\beta+\beta\zeta^\dagger-m\beta|^2
+|h\alpha|^2 + \mbox{X-terms}
\nn\ee
where the cross terms are given by,
\be
\mbox{X-terms}&=& 
(\partial_3\alpha+[\zeta,\alpha], h^\dagger \beta) 
+ (h^\dagger\beta, \partial_3\alpha+[\zeta,\alpha]) \nn\\
&& + (\partial_3\beta+\beta \zeta^\dagger-m\beta, h\alpha)
+(h\alpha, \partial_3\beta+\beta\zeta^\dagger- m \beta) \nn\\
&=&(\partial_3\alpha,  h^\dagger\beta))
+(\alpha, [\zeta^\dagger, h^\dagger \beta])
+(\beta, h(\partial_3\alpha+[\zeta, \alpha])) \nn\\
&&+(\partial_3 \beta, h\alpha) 
+(\beta, h\alpha\zeta -mh\alpha) 
+(\alpha, h^\dagger(\partial_3\beta 
+ \beta \zeta^\dagger-m\beta)) \nn\\
&=&
(\alpha, (-\partial_3h^\dagger+\zeta^\dagger h^\dagger
-h^\dagger m)\beta)
+(\beta, (-\partial_3 h + h\zeta -m h)\alpha)
\nn
\ee
To reach the second and third equalities, 
we have made use of the cyclic property of the trace 
and integrated by parts, respectively. 
The equations in the parentheses are simply the 
transpose of the domain wall 
equations \eqn{vort2} and therefore vanish on the 
background soliton. Since these cross-terms vanish, 
the action of $\Delta^\dagger$ can be written as a 
sum of total squares as in \eqn{nozero}, which 
includes the term $|h\alpha|^2$ and, for $h$ of maximal rank ensures 
that any zero mode of $\Delta^\dagger$ must have  $\alpha=0$. 
To show that the zero mode also requires $\beta=0$ is only slightly 
more involved. The term $|h^\dagger\beta|^2$ can vanish for 
$\beta=-iTh$, where $T$ is a hermitian $N_f\times N_f$ matrix. With 
this ansatz, one can show that the matrix $T$ must solve something 
akin to a lax equation: $\partial_3 T=[m,T]$ which, imposing the 
boundary condition $T=0$ at $x^3=\pm\infty$ requires $T=0$. We conclude 
$\Delta^\dagger$ has no zero modes. 

\para
The number of complex zero modes of $\Delta$ is
defined by ${\cal I}=\lim_{M^2\rightarrow 0}{\cal I}(M^2)$
where the regulated index ${\cal I}(M^2)$ is
\be
{\cal I}(M^2)
={\Tr}\left(\frac{M^2}{\Delta^\dagger\Delta +M^2}\right)
-{\Tr}\left(\frac{M^2}{\Delta\Delta^\dagger
+M^2}\right)
\ee
Clearly ${\cal I}$ counts the number of zero modes 
of $\Delta$ minus the number of zero modes of 
$\Delta^\dagger$. But since, as we have seen, 
the latter operator has vanishing kernel, ${\cal I}$ 
counts the number of parameters of the domain wall 
equations as required. 
Using standard Callias index theorem techniques 
(see for example \cite{erick} for application to 
monopoles, and \cite{keith} for application to domain walls), 
we can rewrite the index as 
\be
{\cal I}(M^2)=-\ft12\left[J(x,M^2)\right]^{x=+\infty}_{x=-\infty}
\nn\ee
where the current $J$ is defined as 
\be
J(x,M^2)
=-\tr\langle x |\ 
\Delta\frac{1}{\Delta^\dagger\Delta+M^2}
+\Delta^\dagger\frac{1}{\Delta\Delta^\dagger +M^2}
\ |x\rangle
\ee
Here $\tr$ denotes the trace over colour and flavour indices, 
as well as over the $2\times 2$ entries of $\Delta$. 

We first need to compute $\Delta\Delta^\dagger$ and 
$\Delta^\dagger\Delta$. In general these are somewhat 
complicated matrices, but our task 
is made a little simpler by the fact that we only 
need to evaluate them in a vacuum state. 
This means that we may set $A=0$ or, in our current 
notation, $\zeta=\zeta^\dagger=\phi^T$. 
We can also insist that derivatives of the background 
field vanish: $\partial\zeta=\partial h=0$. 
Finally, from the F-term condition, we also have that 
$h\zeta=mh$. 
With these constraints, the 
two matrix products may be easily evaluated to give
\be
\left.\Delta^\dagger\Delta\right|_{\rm vacuo}
=\left.\Delta\Delta^\dagger\right|_{\rm vacuo}=
\left(\begin{array}{cc}
-\partial_3^2+\zeta_a^2+h^\dagger h & 0 \\ 0 &
-\partial_3^2+(\zeta_R-m)^2+hh^\dagger\end{array}\right) .
\label{dddagger}\ee
Let's firstly examine the current with 
the expressions we derived above, neglecting 
the trace over flavour indices and the integral over space. 
We have 
\be
&&\tr_2{\Tr}_{N_c}\left[\Delta\frac{1}{\Delta^\dagger\Delta+M^2}
+\Delta^\dagger\frac{1}{\Delta\Delta^\dagger +M^2}
\right]
={\Tr}_{N_c}
\left[(\partial_3-\zeta_a^\dagger)\frac{1}{\Theta_1+M^2}
\right.
\nn\\
&&\left.
+(\partial_3-\zeta_R+m)\frac{1}{\Theta_2+M^2}
%&&\ \ \ \ \ \ \ \ \ \ \ \ \ 
-(\partial_3+\zeta_a)\frac{1}{\Theta_1+M^2}
-(\partial_3+\zeta_R^\dagger-m)\frac{1}{\Theta_2+M^2}\right]
\nn\ee
where, for simplicity of notation, we have defined the operators $\Theta_1$ and $\Theta_2$ arising in
the expression \eqn{dddagger} for $\Delta^\dagger\Delta$ and $\Delta\Delta^\dagger$
\be
\left.\Delta^\dagger\Delta\right|_{\rm vacuo}=\left.\Delta\Delta^\dagger\right|_{\rm vacuo}
=\left(\begin{array}{cc}\Theta_1 & 0 \\ 0 & \Theta_2\end{array}
\right)
\ee
We see that the $\partial$ derivatives in the numerator explicitly cancel, while the adjoint
action $\zeta_a$ vanishes upon taking the trace over the gauge group. Recalling that $\zeta=\zeta^\dagger$
in vacuo, we're left with,
\be
\tr_2{\Tr}_{N_c}\left[\Delta\frac{1}{\Delta^\dagger\Delta+M^2}
+\Delta^\dagger\frac{1}{\Delta\Delta^\dagger +M^2}
\right]=-2{\Tr}_{N_c}\left[(\zeta_R-m)\frac{1}{\Theta_2+M^2}\right]
\ee
Now we can evaluate the differential operator $\partial_3^2$ 
in momentum space to obtain 
%do the integral over the space to get the expression for the current,
\be
J(x,M^2)={\Tr}_{N_c}{\Tr}_{N_f}
\left[\frac{(\zeta_R-m)}{\sqrt{(\zeta_R-m)^2+hh^\dagger+M^2}}\right]
\ee
which, in turn, yields an expression for the number of zero modes:
\be
{\cal I}={\cal I}_{\infty}-{\cal I}_{-\infty}=
-\frac{1}{2}{\Tr}_{N_c}{\Tr}_{N_f}\left[\frac{(\zeta_R-m)}{\sqrt{(\zeta_R-m)^2+hh^\dagger}}
\right]^{+\infty}_{-\infty}
\ee
where ${\cal I}_{\pm\infty}$ denotes the expression in the bracket evaluated at $x^3\rightarrow\pm \infty$.
We are nearly there. All that is left is some algebraic manipulation. Recall from \eqn{vacuum} that
as $x^3\rightarrow +\infty$, we have $\phi={\rm diag}(m_{\sigma(1)},\ldots,m_{\sigma(N_c)})$
and $q^a_{\ i}=v\,\delta^a_{\ \sigma(a)}$.
The combination of $q$ that appears in the expression for the index is
\be
hh^\dagger=q^T(q^T)^\dagger=v^2I_{\infty}
\nn\ee
where the $N_f\times N_f$ matrix $I_{\infty}$ has only non-zero diagonal entries which
are given by $(I_{\infty})_{ii}=1$ if $i\in\Sigma$. We therefore have
\be
-2{\cal I}_{+\infty}&=&{\Tr}_{N_c}{\Tr}_{N_f}\left.\frac{(\phi-m_i)}{\sqrt{(\phi-m_i)^2+q^T(q^T)^\dagger}}\ \right|_{x=+\infty}
\nn\\
&=&\sum_{a=1}^{N_c}\sum_{i=1}^{N_f}\,\frac{(m_{\sigma(a)}-m_i)}{\sqrt{(m_{\sigma(a)}-m_i)^2+v^2I_{\infty}}} \nn\\
&=&\sum_{a=1}^{N_c}\left(\sum_{i\notin\Sigma}{\rm sign}(m_{\sigma(a)}-m_i)+\sum_{i\in\Sigma}
\frac{(m_{\sigma(a)}-m_i)}{\sqrt{(m_{\sigma(a)}-m_i)^2+v^2}}\right) \nn\\
&=&\sum_{a=1}^{N_c}\sum_{i\notin\Sigma}\sign(m_{\sigma(a)}-m_i)+\sum_{a,b=1}^{N_c}
\frac{(m_{\sigma(a)}-m_{\sigma(b)})}{\sqrt{(m_{\sigma(a)}-m_{\sigma(b)})^2+v^2}} \nn\\
&=&\sum_{a=1}^{N_c}\sum_{i\notin\Sigma}\sign(m_{\sigma(a)}-m_i)
\ee
where in the final equality, we have noticed that the term involving sums over $a$ and $b$ vanishes
by anti-symmetry. In fact, this same trick also allows us to rewrite the final term as a sum over all $i$,
rather than the complement of $\Sigma$. So, including the contribution from $x=-\infty$ as well, we
finally get the following expressions for the index
\be
{\cal I}=-\ft12\sum_{i=1}^{N_f}\sum_{a=1}^{N_c}\left(\sign(m_{\sigma(a)}-m_i)-
\sign(m_{\tilde{\sigma}(a)}-m_i)\right)
=-\sum_{a=1}^{N_c}\left(\tilde{\sigma}(a)-\sigma(a)\right)
\label{index}\ee
Finally, we want to massage this expression to write it in terms of the root
vector $\vec{g}$. We'll rewrite the topological charge as $\vec{g}=(p_1,p_2,\ldots,p_{N_f})$
where $p_k=n_k-n_{k-1}\in\{0,1\}$ and 
$\sum_{k=1}^{N_f}p_k=0$. 
Since the vacuum set $\Sigma$ consists of $p_1$ copies of $m_1$ 
and $p_2$ copies of $m_2$, we can write 
$\sum_a(\tilde{\sigma}(a)-{\sigma}(a))=\sum_k kp_k$
which, defining $n_0=n_{N_f}=0$, allows us to rewrite the index as
\be
{\cal I}&=&-\sum_{k=1}^{N_f}kp_k
=-\sum_{k=1}^{N_f}k(n_k-n_{k-1})
= \sum_{a=1}^{N_f-1}n_a
\ee
as advertised.

\subsection*{The Acknowledgments}

Thanks to Maciej Dunajski, Matt Headrick and Paul Townsend for useful 
discussions. N.S. wishes to thank a fruitful collaboration 
with Minoru Eto, Youichi Isozumi, Muneto Nitta, 
Keisuke Ohashi, Kazutoshi Ohta, and Yuji Tachikawa. 
N.S. is supported in part by Grant-in-Aid for Scientific 
Research from the Ministry of Education, Culture, Sports, 
Science and Technology, Japan No.13640269 
and 16028203 for the priority area ``origin of mass''. 
D.T. is supported by the Royal 
Society through a University Research Fellowship.

\end{document}